\documentclass[runningheads]{llncs}
\newcommand{\repeatthanks}{\textsuperscript{\thefootnote}}

\usepackage{graphicx} 
\usepackage{multirow}
\usepackage{listings}
\usepackage{xcolor}
\usepackage{amsmath}
\usepackage[most]{tcolorbox}
\usepackage{booktabs}

\definecolor{darkgreen}{rgb}{0, 0.44, 0.23}
\definecolor{lightgreen}{rgb}{0.25, 0.63, 0.4375}
\definecolor{darkblue}{rgb}{0.05, 0.20, 0.60}
\lstdefinestyle{mystyle}{
    basicstyle=\footnotesize\ttfamily,
    breaklines=true,
    captionpos=b,
    numbers=left,
    keywords={}
    xleftmargin=\parindent,
    keywordstyle=\color{darkgreen},
    tabsize=2,
    emph={Type, Import, Function, Table, Memory, Global, Export, Start, Elem, Code, Data, Local, Instruction},
    emphstyle=\color{darkblue},
    showstringspaces=false,
    keywordstyle=\color{magenta},
    otherkeywords={insert, select, update, delete, modifyFuncInstr, appendImportInstr, InternalFunction, HookFunction},
    backgroundcolor=\color{white},
    deletekeywords={len},
    escapechar={|}, 
}
\newcommand{\code}[1]{{\ttfamily #1}}
\newcommand{\framework}{\textsc{BREWasm}}

\begin{document}

\title{A General Static Binary Rewriting Framework for WebAssembly}

\titlerunning{Static Binary Rewriting for WebAssembly}

\author{Shangtong Cao\inst{1}\thanks{The first two authors contribute equally}\and
Ningyu He\inst{2}\repeatthanks\and
Yao Guo\inst{2}\and
Haoyu Wang\inst{3}}
\authorrunning{S. Cao and N. He et al.}

\institute{Beijing University of Posts and Telecommunications\and
Peking University\and
Huazhong University of Science and Technology}

\maketitle

\begin{abstract}
Binary rewriting is a widely adopted technique in software analysis. WebAssembly (Wasm), as an emerging bytecode format, has attracted great attention from our community. Unfortunately, there is no general-purpose binary rewriting framework for Wasm, and existing effort on Wasm binary modification is error-prone and tedious. In this paper, we present {\framework}, the first general purpose static binary rewriting framework for Wasm, which has addressed inherent challenges of Wasm rewriting including high complicated binary structure, strict static syntax verification, and coupling among sections. We perform extensive evaluation on diverse Wasm applications to show the efficiency, correctness and effectiveness of {\framework}. We further show the promising direction of implementing a diverse set of binary rewriting tasks based on {\framework} in an effortless and user-friendly manner. 

\end{abstract}

\section{Introduction}
\label{sec:intro}
WebAssembly (Wasm)~\cite{wasm}, endorsed by Internet giants like Google and Mozilla, is an assembly-like stack-based low-level language, aiming to execute at native speed.
Portability of Wasm is achieved by the ability of being a compiling target for mainstream high-level programming languages, e.g., C/C++~\cite{c/c++}, Go~\cite{Go}, and Rust~\cite{Rust}.
Lots of resource-consumed and -sensitive software have been compiled to Wasm binaries and embedded in browsers~\cite{scnerios}, like 3D graphic engines, Scientific operations, Computer-aided design, and multimedia encoders and decoders.
Beyond the browser, Wasm is moving towards a much wider spectrum of domains, including IoT \cite{iot}, serverless computing \cite{Serverless}, edge computing \cite{edge}, blockchain and Web 3.0 \cite{eosio}, etc.

The rising of Wasm has attracted massive attention from our research community. As an emerging instruction format, our fellow researchers have invest huge effort into Wasm binary analysis, including static analysis for vulnerability detection~\cite{static-analysis1}, dynamic analysis based on program instrumentation~\cite{lehmann2019wasabi}, Wasm binary transformation~\cite{binary-mutate1}, and binary optimization~\cite{binary-optimization}, etc. Most existing studies, rely on Wasm binary rewriting to implement their functionalities, more or less. For example, Wasabi~\cite{lehmann2019wasabi} is a dynamic analysis framework of Wasm that can dynamically obtain the runtime information of target binary via instrumenting Wasm instructions. However, due to the case-specific demands of existing work, researchers need to implement a specific set of rewriting rules from scratch, or even manually modify Wasm binaries, which is error-prone and tedious. We, therefore, argue that a general purpose rewriting framework is necessary to facilitate the research on Wasm binaries.

Binary rewriting is a general technique to modify existing executable programs, which is a well-studied direction for native binaries~\cite{Alto,SASI,Diablo,Egalito,McSema}. Unfortunately, there currently is no general-purpose binary rewriting framework for Wasm. Implementing such a rewriting framework is challenging.
First, Wasm is complicated in both semantics and syntax. There are 11 kinds of valid sections defined in Wasm, and each section is composed of vectors, each of which is further composed of several attributes, which can be further wrapped to make the syntax super complex. However, as a user-friendly general rewriting framework, it should handle all of these complexity in a concise way, which is a natural contradiction.
Second, even modifying a tiny functionality in Wasm usually requires updating several sections accordingly. For example, if a user wants to insert a new function to the Wasm binary, he has to update several sections, including the type, the function, and the code sections. Moreover, if the added function can be taken as the callee of a function pointer, the table section and the elem section should also be updated accordingly. Manually update all sections is tedious and error-prone.
Besides, Wasm enforces strict verification before executing, and any violations against the Wasm syntax during rewriting Wasm binaries will invalidate them. An invalid Wasm binary cannot be loaded and executed normally. However, any tiny operation provided by the binary rewriting framework would break the verification, which should be paid special attentions during rewriting.

\textbf{This Work.}
In this paper, we implement {\framework}, a general-purpose binary rewriting framework for Wasm. {\framework} is composed of four components: \textit{Wasm Parser}, \textit{section rewriter}, \textit{semantics rewriter}, and \textit{Wasm Encoder}.
Specifically, the Wasm parser and encoder are implemented based on our abstraction of the Wasm binary, which is represented as a formal format, with a list of objects. Based on these abstracted objects, the section rewriter is able to conduct rewriting, e.g., inserting/deleting a new object or modifying attributes of existing objects. It packs these fine-grained rewriting functions into APIs.
The semantics rewriter further combines these fine-grained APIs and offers another set of high-level APIs, where each of them possesses rich semantics, like inserting a function, and append a piece of linear memory.
Through these exposed APIs, Wasm binaries can be arbitrarily modified for any purposes without considering the underlying complexity of semantics and syntax.

Based on benchmarks consisting of representative Wasm binaries, the evaluation results show the efficiency, correctness, effectiveness, and real-world usability of {\framework}.
Specially, it is practical to achieve various kinds of complicated Wasm binary rewriting tasks by combing the APIs provided by {\framework}, including binary instrumentation, binary hardening and mutation-based fuzzing. Comparing with the cumbersome implementation of the specific tasks, the work built on {\framework} is effortless and user-friendly.

\textbf{Our contribution} can be summarized as follows:
\begin{itemize}
\item To the best of our knowledge, we have implemented the first general purpose Wasm binary rewriting framework, which offers more than 31 semantic APIs that are summarized from real-world usage scenarios.
\item We perform extensive evaluation on diverse Wasm applications to show the efficiency, correctness and effectiveness of {\framework}.
\item We show that it is useful, effortless and user-friendly to implement a diverse set of binary rewriting tasks based on {\framework}.

\end{itemize}

To boost further research on Wasm binary rewriting, we will release {\framework} to the research community.

\section{Background}
\label{sec:background}
In this section, we will briefly illustrate some basic concepts of Wasm and binary rewriting.

\subsection{Webassembly Binary}
\label{sec:backgroud:wasm}
WebAssembly (Wasm) is an emerging cross-platform low-level language that can be compiled by multiple mainstream high-level programming languages~\cite{multi-languages}, e.g., C/C++~\cite{c/c++}, Rust~\cite{Rust}, and Go~\cite{Go}.
Wasm is designed to be effective and compact. It can achieve nearly native code speed in performance with a sufficiently small size (100KB to 1MB for an ordinary Wasm binary~\cite{binary-size}).
In addition, some official auxiliary tools are proposed to facilitate the development of Wasm.
For example, \textit{wasm2wat}~\cite{wasm2wat} can translate a Wasm binary into WebAssembly Text Format (WAT for short), and \textit{wasm-validate}~\cite{wasm-validate} can be invoked to validate the correctness of the syntax of a Wasm binary.

\begin{figure}[!t] 
\centering 
\includegraphics[width=0.8\columnwidth]{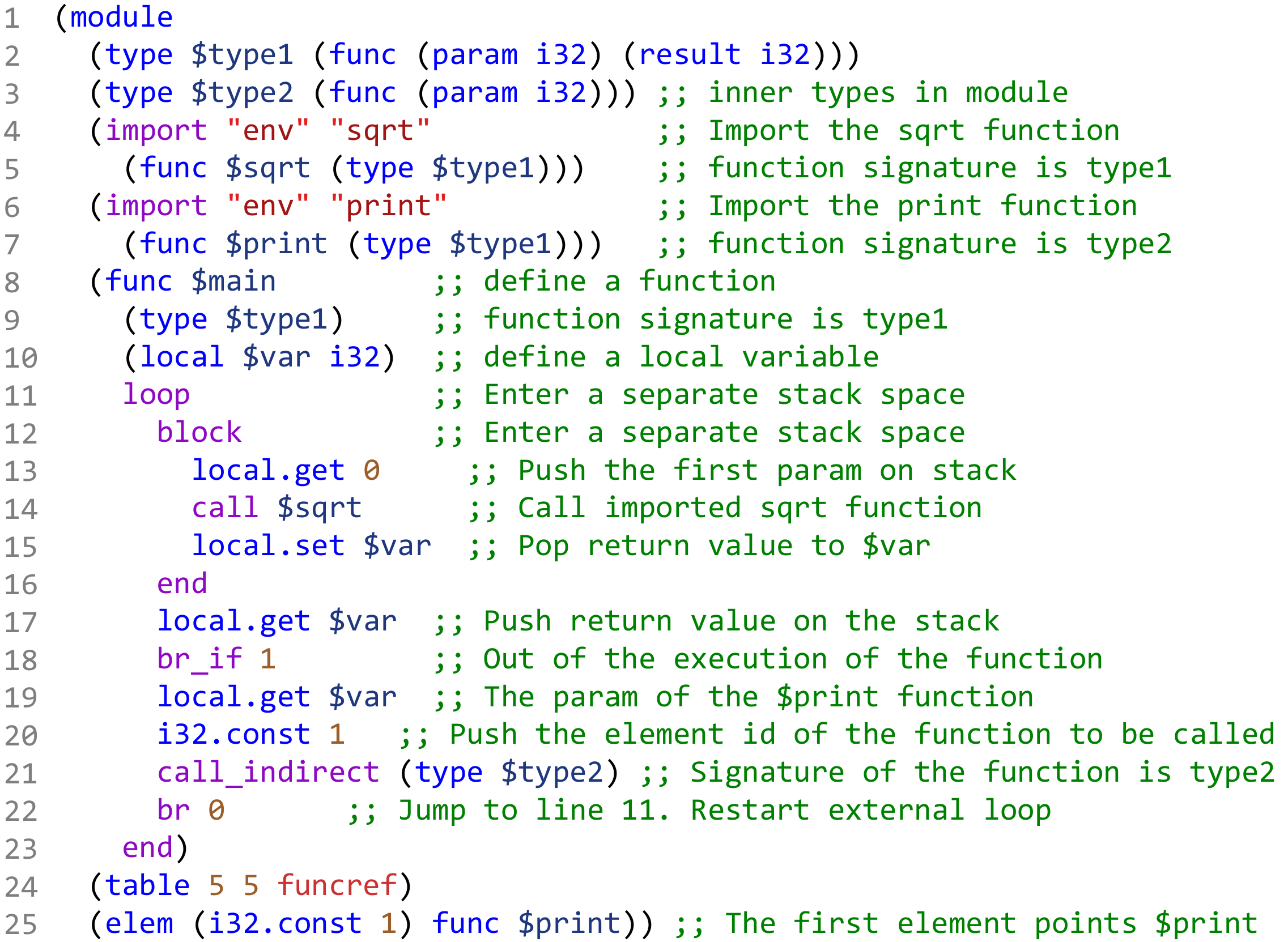} 
\caption{A code snippet of WebAssembly in WAT text format.} 
\label{fig:wasm-snippet} 
\end{figure}

Each Wasm binary is composed of sections with different functionalities.
Specifically, in a Wasm binary, functions are implemented in the \textit{\textbf{code section}}, and their signatures are declared in the \textit{\textbf{type section}}. The \textit{\textbf{function section}} maintains a mapping from the index of each function to the index of its corresponding type.
Functions can also be imported from environment through the \textit{\textbf{import section}}, and be exported via the \textit{\textbf{export function}}.
Except for their independent contexts, data stored in the \textit{\textbf{global section}}, \textit{\textbf{data section}}, and \textit{\textbf{memory section}} can be accessed arbitrarily.
To implement function pointers, Wasm designs an \textit{indirect call mechanism}, where all possible callees have to be declared in the \textit{\textbf{table section}} and \textit{\textbf{element section}}.
According to the compiling options, debugging information or third-party extensions will be stored in the \textit{\textbf{custom section}}, which has no effect on execution.

\begin{figure}[!t] 
\centering  
\includegraphics[width=0.8\columnwidth]{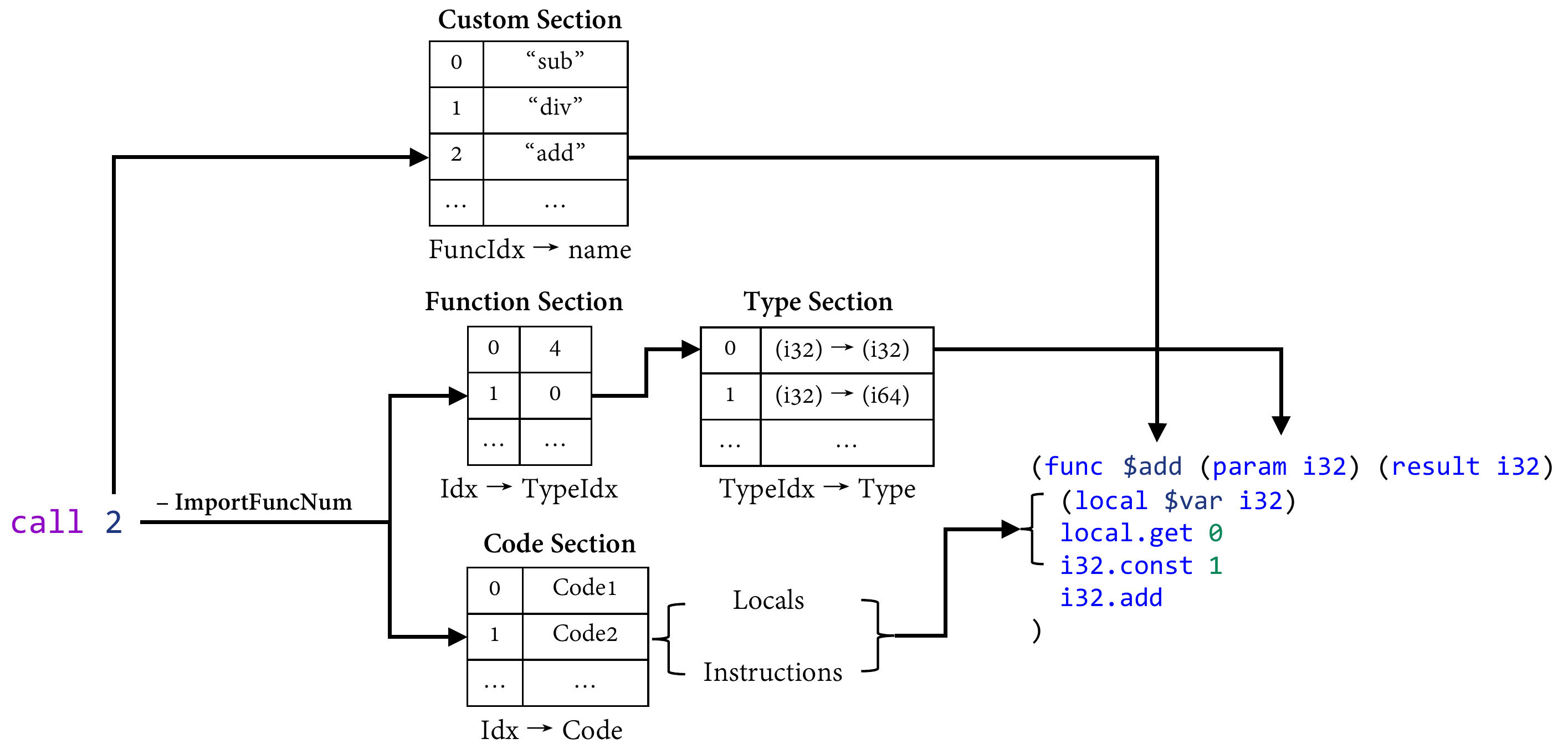} 
\caption{Function indexing is achieved by coupling several sections} 
\label{fig:index-func} 
\end{figure}

Sections can be further divided into \textit{vectors}, the smallest unit that declares a functionality.
For example, Fig.~\ref{fig:wasm-snippet} shows a code snippet of Wasm binary (shown in WAT text format). As we can see, L2\footnote{The second line, denoted by L2. We adopt such notations in the following.} is a vector belonging to \textit{\textbf{type section}}, which declares a function signature.
A vector may have many attributes, e.g., the vector at L2 consists of its index (\texttt{\$type1}), parameters type (\texttt{i32}), and the return value type (\texttt{i32}).
In Wasm, however, \textit{a single semantics is often achieved by coupling several sections}.
Taking indexing a function as an example, which is shown in Fig.~\ref{fig:index-func}.
Once an internal function, indexed by 2 in this example, is invoked by a \texttt{call} instruction, its readable name can be indexed via the \textit{custom section}. To obtain its signature, we need a two-layer translation through the \textit{function section} and the \textit{type section}. Its implementation can be accessed only via the \textit{code section}.
Note that, all imported functions are located in front of normal functions, thus we need to subtract the number of imported functions, which is 1 in this example, to obtain its real index when indexing in the function section and the code section.

\subsection{Binary Rewriting}
Binary rewriting\footnote{In this work, the \textit{binary rewriting} specifically refers to static binary rewriting.} refers to the process of taking a binary as input, rewriting various parts of it, and generating another binary that is properly formatted. The semantics of the rewritten program depend on the rewriting purpose and strategy. This technique has been widely adopted in the software analysis direction, e.g., program instrumentation~\cite{binary-instrumentation1,binary-instrumentation2,binary-instrumentation3,binary-insturmentation4}, binary enhancement~\cite{binary-harden1,binary-harden2,binary-harden3}, and program transformation~\cite{binary-mutate1,binary-mutate2,binary-mutate3}, etc.

Currently, some work specifically conducts rewriting against Wasm binaries.
For example, Wasabi~\cite{lehmann2019wasabi} is a dynamic analysis framework of Wasm. It can dynamically obtain the runtime information of target binary via instrumenting Wasm instructions.
Fuzzm~\cite{Lehmann:2021aa} statically inserts stack canaries into linear memory. To this end, it can identify memory bugs by fuzzing the instrumented Wasm binary.
SEISMIC~\cite{SEISMIC} also conducts instrumentation against Wasm binaries. However, the process specifically aims at instructions commonly used in cryptomining algorithms, e.g., \texttt{i32.add} and \texttt{i32.shl}. Through running the instrumented program, the analyzer can determine whether the target is a malicious mining program according to its profiling results.
At last, Wasm-mutate~\cite{wasm-mutate} is a binary mutation tool. It integrates many different strategies for performing mutation on Wasm binaries, e.g., deliberately inserting functions or a piece of memory.
All of these tools rely on binary rewriting to implement the core functionalities. However, due to their case-specific demands, developers need to implement a specific set of rewriting rules from scratch, or even manually modify Wasm binaries, which is error prone and tedious. We, therefore, argue that a general purpose rewriting framework is necessary to facilitate the research on Wasm binaries.

\section{{\framework}}
\label{sec:overview}
In this section, we first overview the architecture of {\framework}, and then detail the challenges in designing and implementing {\framework}.

\begin{figure}[!t] 
\centering  
\includegraphics[width=0.8\columnwidth]{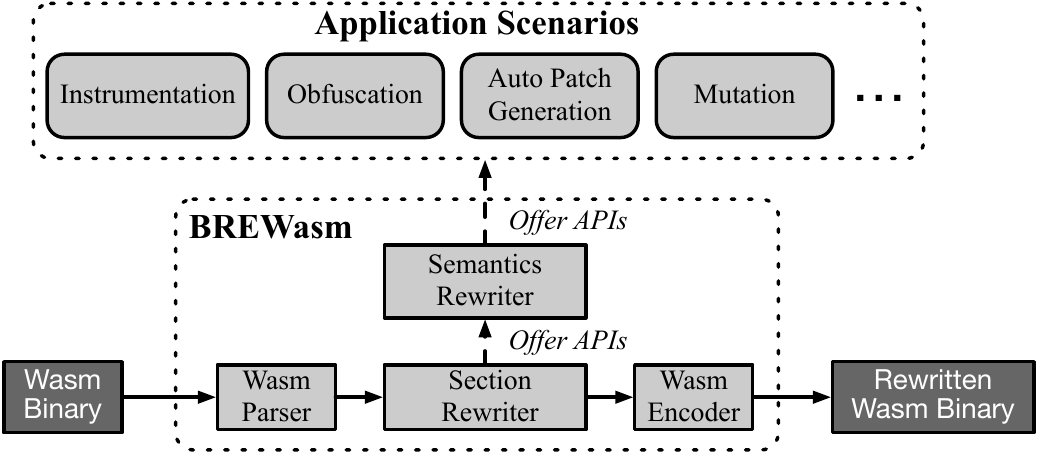} 
\caption{The architecture and workflow of {\framework}.} 
\label{fig:chaos} 
\end{figure}

\subsection{Overview}
To the best of our knowledge, we have implemented the first general rewriting framework against Wasm binaries, dubbed as {\framework}, whose architecture and workflow are shown in Fig.~\ref{fig:chaos}.
As we can see, {\framework} is composed of four components: \textit{Wasm Parser}, \textit{section rewriter}, \textit{semantics rewriter}, and \textit{Wasm Encoder}.
Specifically, the Wasm parser takes a Wasm binary as input, and parses it as a list of objects.
Based on these objects, the section rewriter is able to conduct rewriting, e.g., inserting/deleting a new object or modifying attributes of existing objects. It packs these fine-grained rewriting functions into APIs.
The semantics rewriter combines these fine-grained APIs and offers another set of high-level APIs, where each of them possesses rich semantics, like inserting a function, and append a piece of linear memory.
Through these exposed APIs, users can rewrite Wasm binaries for different goals, e.g., obfuscation, instrumentation, or patch generation on vulnerabilities.
Finally, these updated objects will be encoded into a valid Wasm binary.
The implementation of {\framework} is detailed in \S\ref{sec:approach}, and some usage scenarios will be depicted in \S\ref{sec:case-study}.

\subsection{Challenges}
\label{sec:overview:challenges}
Implementing a general rewriting framework against binaries is often challenging. Because binaries are highly structured and have little semantic information to guide the rewriting process.
As for Wasm binaries, as we introduced in \S\ref{sec:backgroud:wasm}, although Wasm supports inter-conversion between binary and text formats, it is still infeasible to directly modify its text format for rewriting purposes. This can be attributed to three points, i.e., \textit{complicated format}, \textit{strict static verification}, and \textit{coupling among sections}.
We will detail these three challenges in the following.

\noindent
\textbf{C1: Format Complexity.}
Conducting efficient and effective static binary rewriting is strongly correlated to the complexity of the rewritten binary. From Fig.~\ref{fig:wasm-snippet}, we can see that Wasm is highly complex.
In nutshell, the complexity of Wasm is reflected in two aspects.
On the one hand, Wasm has complicated \textit{\textbf{semantics}}.
As we introduced in \S\ref{sec:backgroud:wasm}, there are 11 valid sections defined under the current specification~\cite{wasm-spec}. Each section is composed of vectors, each of which is further composed of several attributes. As an assembly-like language, each attribute is indispensable and corresponds to a specific and unique meaning.
On the other hand, Wasm is also complicated in \textit{\textbf{syntax}}.
Wasm has a highly structured syntax, which also varies in sections.
For example, for a vector in type section (see L2 in Fig.~\ref{fig:wasm-snippet}), the attributes \texttt{param} and \texttt{result} are wrapped by a \texttt{func}, which is further wrapped by the corresponding \texttt{type}.
The same situation also plays for other sections.
Therefore, the semantic and syntactic complexity of Wasm makes it exceptionally challenging to implement a general rewriting framework.

\noindent
\textbf{Our Solution:}
In order to enhance readability and facilitate the following rewriting process, we implement a parser to translate the given Wasm binary into an array of \textit{objects}, each of which is composed of several \textit{attributes}.
During the parsing process, we also omit some auxiliary strings, like the \texttt{type} and the \texttt{func} at L2 of Fig.~\ref{fig:wasm-snippet}. Therefore, each highly structured and nested vectors will be translated into an object with several side-by-side attributes according to its semantics.
Binary rewriting can be easily performed by modifying objects.
Meanwhile, we also implement an encoder to conduct the opposite process, i.e., translating objects into a Wasm binary.
Please see \S\ref{sec:approach:parser-encoder} for more details.

\noindent
\textbf{C2: Static Verification.}
Each Wasm binary will be thoroughly and strictly verified statically before executing~\cite{bring-web-with-wasm,static-validate}. Such a static verification performs on several aspects.
For example, sections are composed of vectors, each of which is indexed by an index. Thus, if a user inserts/deletes a vector, he has to update vectors of the whole section to ensure the continuity of indices of vectors.
Moreover, as we mentioned in \textbf{C1}, the encoder has to reassemble objects and complete the auxiliary strings that are discarded in the parsing process. Any negligence will invalidate the rewritten Wasm binary.

\noindent
\textbf{Our Solution:}
To solve this problem, we implement a fixer that will be automatically invoked after each time of invoking the APIs exposed by the section rewriter.
Specifically, the fixer is mainly responsible for repairing the incontinuity for indices of the rewritten section.
It can also fix some context-aware errors, like increasing the limitation (if necessary) to hold a newly inserted memory.
In addition, after the encoding process, the official syntactic checker, wasm-validate, will be invoked to examine the validity of the rewritten Wasm binary.
Please refer to \S\ref{sec:approach:section-rewriter}.

\noindent
\textbf{C3: Sections Coupling.}
As we mentioned in \S\ref{sec:backgroud:wasm}, some functionalities should be achieved by combining multiple sections.
For example, if we add an extra function in a Wasm binary, except for inserting its implementation in the code section, we have to modify the function section and the type section to declare its signature.
Moreover, if the added function can be taken as the callee of a function pointer, the table section and the elem section should also be updated accordingly.
Manually update all sections is tedious and error-prone.
Such a section coupling problem raises another challenge for the user to achieve his intended goal.

\noindent
\textbf{Our Solution:}
To address the sections coupling problem, we abstract the coupling between sections into a set of Wasm program semantics, i.e., global variable, import and export, linear memory, function and custom content.
Based on these five semantics, we expose a set of APIs, e.g., \texttt{insertInternalFunc}, which takes a function's body and signature as inputs.
Inside the API, we determine if the signature already exists, and insert a new one if necessary. Then, we will insert its declaration and implementation in the function section and the code section, respectively.
It is worth noting that if any reference relation goes wrong due to indices mismatch, like the index of callee of a \texttt{call} is incremented by 1 due to inserting a function, these reference relations will be repaired automatically after each time of invoking APIs of the semantics rewriter.
Please refer to \S\ref{sec:approach:semantics} for more details.

\section{Approach}
\label{sec:approach}
In this section, we will introduce the technical details of components of {\framework}, and how we address the aforementioned challenges.

\begin{figure}[!t] 
\centering
\includegraphics[width=0.8\columnwidth]{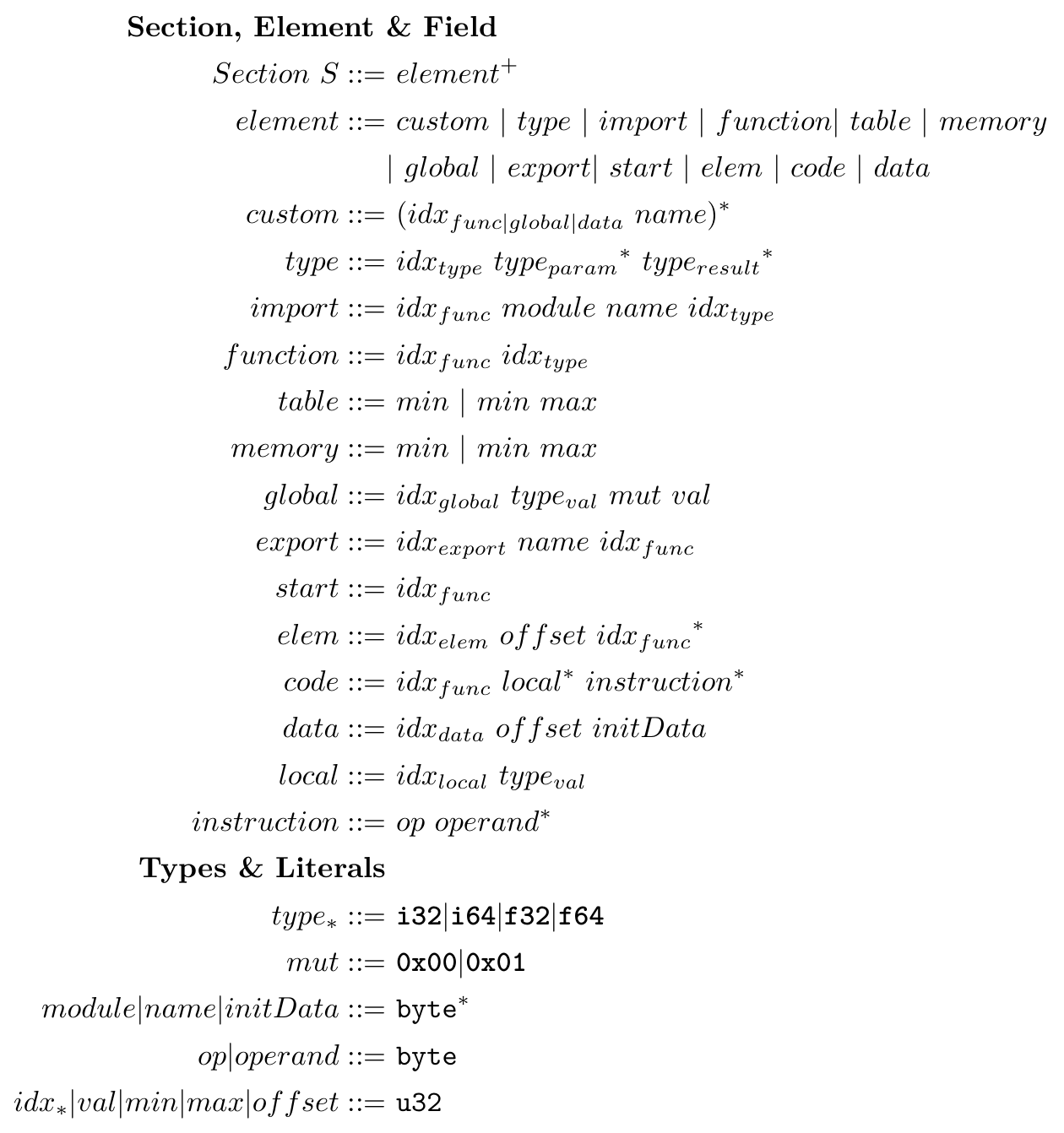}
\caption{Formal definition of sections, elements and fields in Wasm.} 
\label{fig:sec-elem-field} 
\end{figure}

\subsection{Wasm Parser \& Wasm Encoder}
\label{sec:approach:parser-encoder}
Wasm has a highly structured and complicated format. Specifically, a Wasm binary is composed of \textit{sections}, which is a vector of \textit{elements}. Further, an element consists of several \textit{fields} according to the section where it locates.
As we mentioned in \textbf{C1}, to facilitate the following rewriting process, we hide unnecessary and verbose details and translate the Wasm binary into a semantically equivalent format.
Referring to the official Wasm specification, we formally defined the relationships among sections, elements, and fields as shown in Fig.~\ref{fig:sec-elem-field}.

As we can see, each section is composed of a list of elements with the same name, where elements are composed of several fields.
Specifically, each \textit{custom} element is composed of \textit{index-name} pairs. These pairs can be parsed as debugging information for different purposes, like keeping readable names of functions, global variables, and a piece of data.
Moreover, the \textit{type} element consists three fields, indicating the function signature ${type_{param}}^*\rightarrow {type_{result}}^*$ is declared by the $idx_{type}$-th type element.
The definitions of \textit{import} element and \textit{export} element are similar. An \textit{import} element indicates the $idx_{func}$-th function with type declared by $idx_{type}$ is imported from $modulename$ and named as $name$, while an \textit{export} element refers to the $idx_{func}$-th function is exported as $name$, that can be invoked by the environment.
Note that, the \textit{function} element only declares a function's index and its signature, where the implementations are defined by the \textit{code} element.
The $min$ and $max$ defined in \textit{memory} elements jointly limit the available size of the linear memory, and \textit{data} element declares that the initial value ($initData$) of the $idx_{data}$-th linear memory starts from the designated $offset$.
Similarly, \textit{table} elements and \textit{elem} elements share the pattern, but ${idx_{func}}^*$ refers to callee indices for \texttt{call\_indirect} instructions.
Finally, a \textit{global} element declares its value as $val$ with type of $type_{val}$, where $mut$ indicates whether its value can be updated by instructions.
Two extra terms $local$ and $instruction$ are defined which are nestedly adopted in \textit{code} elements.

For each section, we have defined a class with its fields as attributes. Each element is an object of the corresponding class.
To this end, the Wasm parser is able to translate a Wasm binary into a list of objects.
For example, Listing~\ref{lst:parsed-objects} illustrates the parsed objects of the Wasm binary in Fig.~\ref{fig:wasm-snippet}.
\begin{lstlisting}[caption={Parsed objects of the Wasm binary of Fig.~\ref{fig:wasm-snippet}}, label={lst:parsed-objects}]
parsedWasm = [
    Type(0, ["i32"], ["i32"]),
    Type(1, ["i32"], []),
    Import(0, "env", "sqrt", 0),
    Import(1, "env", "print", 0),
    Function(2, 0),
    # omit following instructions
    Code(2, [Local(0, "i32")], [Instruction("0x03", []), ...]),
    Table(5, 5),
    Elem(0, 1, [1])
]
\end{lstlisting}

As we can see from Listing~\ref{lst:parsed-objects}, it translates each elements and packs them into their corresponding objects.
The field names are hidden, but we can obtain the corresponding value according to the definition in Fig.~\ref{fig:sec-elem-field}.
For example, L8 is a \textit{code} element, its first $2$ indicates that it corresponds to the implementation of the second function. Its type is indexed by $0$ (defined by the function element at L6), i.e., \texttt{i32} $\rightarrow$ \texttt{i32} (declared at L2).
The second field of the code element is a list of \textit{local} objects, each of which declares the type and the value of local variables.
Similarly, the third field declares all its instructions, which are wrapped by \texttt{Instruction} objects.
The first instruction has an \textit{op} valued as \texttt{0x03} and an empty \textit{operand}, corresponding to the first \texttt{loop} instruction at L11 in Fig.~\ref{fig:wasm-snippet}.
Through the opposite direction, the Wasm encoder can translate and reassemble these objects into a Wasm binary in a lossless way.

\subsection{Section Rewriter}
\label{sec:approach:section-rewriter}
The section rewriter plays a vital role for {\framework}.
It provides APIs that allow users to manipulate sections on fine-grained level, which is composed of four basic operations: \textit{select}, \textit{insert}, \textit{delete}, and \textit{update}.
Through combining these four operations, users are able to manipulate any element or field we mentioned in \S\ref{sec:approach:parser-encoder}.
The syntax for these operations is formally expressed as follows:
\begin{align*}
\ \ \ \ \ \ \ \ select :&\ element_{template} \rightarrow element^*\\
insert :&\ element^* \times element_{new} \rightarrow \textbf{\texttt{true|false}}\\
delete :&\ element^* \rightarrow \textbf{\texttt{true|false}}\\
update :&\ field \times field_{new} \rightarrow \textbf{\texttt{true|false}}
\end{align*}

Specifically, within the context after parsing the given Wasm binary, the \textit{select} takes an element template ($element_{template}$) to filter out all elements conforming to the $element_{template}$. Note that the wild card, an understrike character, is allowed when designating a field.
For example, $\texttt{select(Type(\_, \_, [`i32']))}$ will return all \textit{type} elements that return a single \texttt{i32} without considering their arguments.
To this end, the object at L2 in Listing~\ref{lst:parsed-objects} instead of the one at L3 will be returned.
Based on the selected results, the \textit{insert} and \textit{delete} can be conducted to insert a new element ($element_{new}$) after the designated one(s) and delete the given elements, respectively.
Take a concrete situation as an instance.
In the Listing~\ref{lst:parsed-objects}, if a user wants to delete the first type element and insert a new one, with arguments as \texttt{i64} and returns as \texttt{i32}, he can write the following two statements:
\begin{align*}
\ \ \ \ \ &\ \ \ \ \ \ \ \ \ \small\texttt{delete(select(Type(0, [`i32'], [`i32'])))}\\
&\small\texttt{insert(select(Type(\_, \_, \_))[-1], Type(\_, [`i64'], [`i32']))}
\end{align*}
, where the first statement deletes the object at L2 by an exact match, and the second statement inserts a new type element after the last one.
Moreover, the \textit{update} can be used to modify a field by a new value ($field_{new}$).
Field values can be retrieved by a dot operator, like getting values of an attribute in an object.
For example, the user intends to modify the returns as \texttt{i64} on the just inserted type element. He can invoke the following statement:
\begin{align*}
\ \ &\small\texttt{update(select(Type(2,[`i64'],[`i32'])).resultType,[`i64'])}
\end{align*}
, where the field $type_{result}$ is accessed by a dot operator with an identical name.

\begin{table*}[tbp]
\centering
\caption{Representative APIs exposed by the semantics rewriter on 5 abstracted semantics.}
\resizebox{\linewidth}{!}{%
\begin{tabular}{p{0.5in}p{0.5in}p{1.85in}p{3in}}
\hline
\textbf{Semantic} &
  \textbf{Sections} &
  \textbf{Representative API(s)} &
  \textbf{Explanations} \\ \hline
\textbf{\begin{tabular}[c]{@{}l@{}}Global\\ Variables\end{tabular}} &
  Global &
  \begin{tabular}[c]{@{}l@{}}\code{insertGlobalVariable}\\ \ \ \ \textit{idx}\ :\ \texttt{u32}\\ \ \ \ \textit{valType}\ :\ \textbf{\texttt{i32}}$|$\textbf{\texttt{i64}}$|$\textbf{\texttt{f32}}$|$\textbf{\texttt{f64}}\\ \ \ \ \textit{mut}\ :\ \textbf{\texttt{0x00}}$|$\textbf{\texttt{0x01}}\\ \ \ \ \textit{initValue}\ :\ \texttt{u32}\end{tabular} &
  \begin{tabular}[c]{@{}l@{}}
  \specialrule{0em}{0.4pt}{0.4pt}
    \begin{minipage}{\linewidth}
      \includegraphics[width=\linewidth, height=5mm]{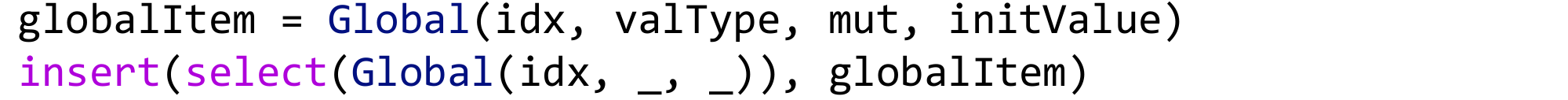}
    \end{minipage}
  \end{tabular} \\ \hline
\multirow{5}{*}{\textbf{\begin{tabular}[c]{@{}l@{}}Import \\ \quad\&\\ Export\end{tabular}}} &
  \multirow{5}{*}{\begin{tabular}[c]{@{}l@{}}Type\\ Import\\ Export\end{tabular}} &
  \begin{tabular}[c]{@{}l@{}}\code{insertImportFunction}\\ \ \ \ \textit{idx}\ :\ \texttt{u32}\\ \ \ \ \textit{moduleName}\ :\ \texttt{byte}$^*$\\ \ \ \ \textit{funcName}\ :\ \texttt{byte}$^*$\\ \ \ \ \textit{paramsType}\ :\ (\textbf{\texttt{i32}}$|$\textbf{\texttt{i64}}$|$\textbf{\texttt{f32}}$|$\textbf{\texttt{f64}})*\\ \ \ \ \textit{resultsType}\ :\ (\textbf{\texttt{i32}}$|$\textbf{\texttt{i64}}$|$\textbf{\texttt{f32}}$|$\textbf{\texttt{f64}})*\end{tabular} &
  \begin{tabular}[c]{@{}l@{}}
     \specialrule{0em}{0.4pt}{0.4pt}
    \begin{minipage}{\linewidth}
      \includegraphics[width=\linewidth, height=32.5mm]{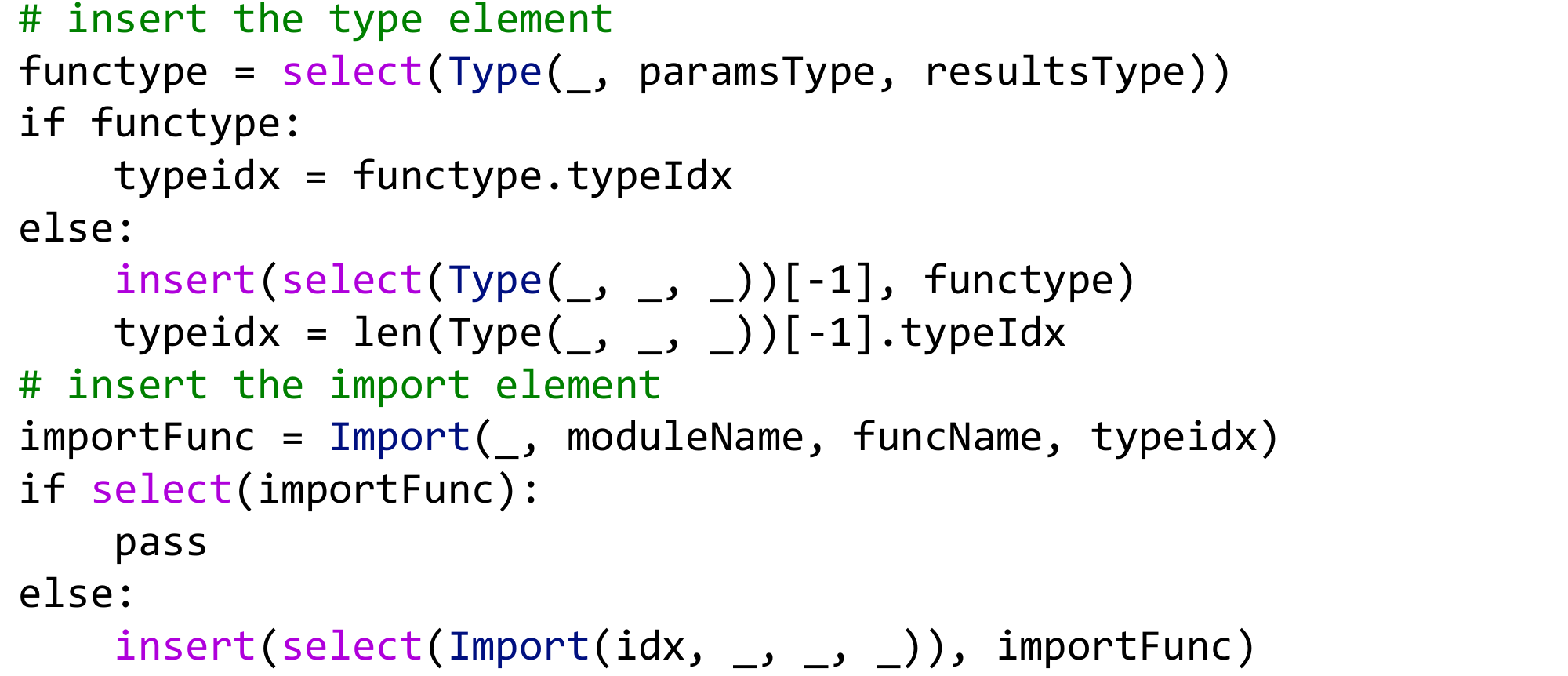}
    \end{minipage}
  \end{tabular} \\ \cline{3-4} 
 &
   &
  \begin{tabular}[c]{@{}l@{}}\code{insertExportFunction}\\ \ \ \ \textit{idx}\ :\ \texttt{u32}\\ \ \ \ \textit{funcName}\ :\ \texttt{byte}$^*$ \\ \ \ \ \textit{funcidx}\ :\ \texttt{u32}\end{tabular} &
  \begin{tabular}[c]{@{}l@{}}
     \specialrule{0em}{0.4pt}{0.4pt}
    \begin{minipage}{\linewidth}
      \includegraphics[width=\linewidth, height=15mm]{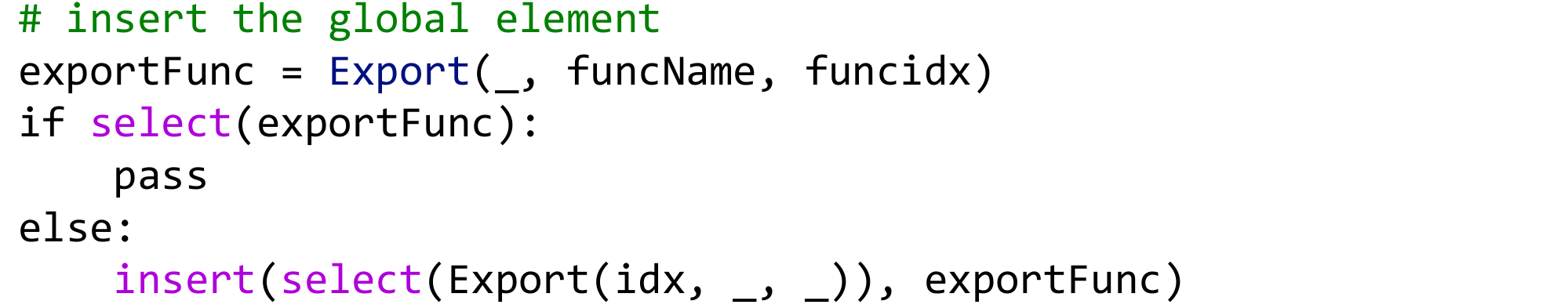}
    \end{minipage}
  \end{tabular} \\ \hline
  
\multirow{7}{*}{\textbf{\begin{tabular}[c]{@{}l@{}}Linear\\ Memory\end{tabular}}} &
  \multirow{7}{*}{\begin{tabular}[c]{@{}l@{}}Memory\\ Data\end{tabular}} &
  \begin{tabular}[c]{@{}l@{}}\code{appendLinearMemory}\\ \ \ \ \textit{pageNum}\ :\ \texttt{u32}\end{tabular} &
  \begin{tabular}[c]{@{}l@{}}
     \specialrule{0em}{0.4pt}{0.4pt}
    \begin{minipage}{\linewidth}
      \includegraphics[width=\linewidth, height=7.5mm]{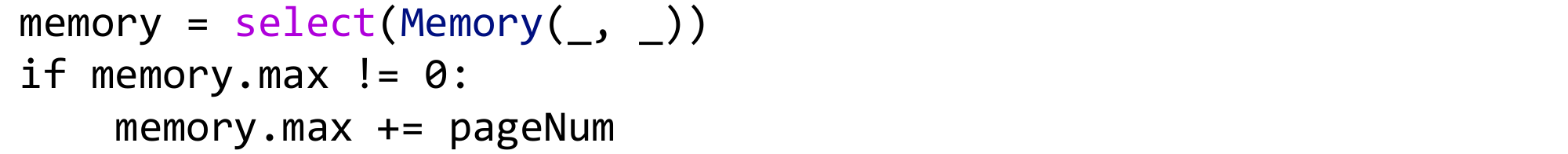}
    \end{minipage}
  \end{tabular} \\ \cline{3-4} 
 &
   &
  \begin{tabular}[c]{@{}l@{}}\code{modifyLinearMemory}\\ \ \ \ \textit{offset}\ :\ \texttt{u32} \\ \ \ \ \textit{bytes}\ :\ \texttt{byte}$^*$ \end{tabular} &
  \begin{tabular}[c]{@{}l@{}}
      \specialrule{0em}{0.4pt}{0.4pt}
    \begin{minipage}{\linewidth}
      \includegraphics[width=\linewidth, height=27.5mm]{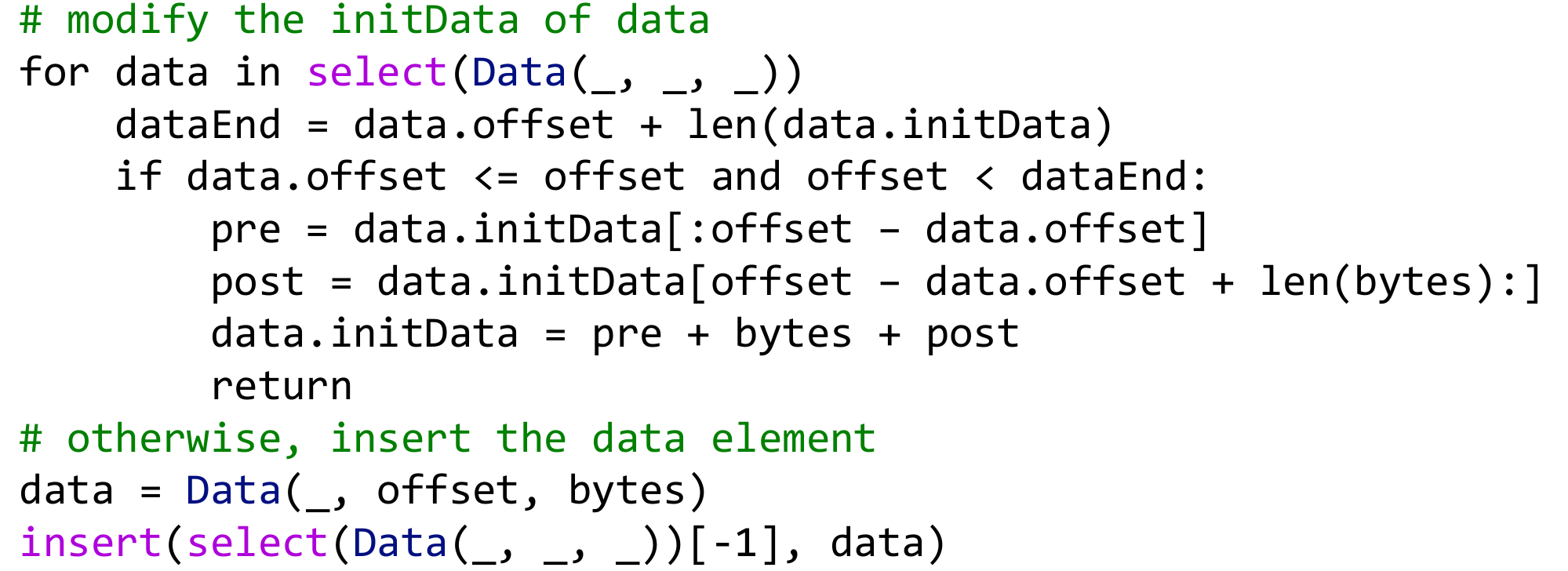}
    \end{minipage}
  \end{tabular} \\ \hline
\multirow{16}{*}{\textbf{Function}} &
  \multirow{16}{*}{\begin{tabular}[c]{@{}l@{}}Type\\ Function\\ Code\\ Start\\ Table\\ Element\end{tabular}} &
  \begin{tabular}[c]{@{}l@{}}\code{insertInternalFunction}\\ \ \ \ \textit{funcidx}\ :\ \texttt{u32} \\ \ \ \ \textit{paramsType}\ :\ (\textbf{\texttt{i32}}$|$\textbf{\texttt{i64}}$|$\textbf{\texttt{f32}}$|$\textbf{\texttt{f64}})*\\ \ \ \ \textit{resultsType}\ :\ (\textbf{\texttt{i32}}$|$\textbf{\texttt{i64}}$|$\textbf{\texttt{f32}}$|$\textbf{\texttt{f64}})*\\ \ \ \ \textit{locals}\ :\ \textit{local}*\\ \ \ \ \textit{funcBody}\ :\ \textit{instruction}*\end{tabular} &
  \begin{tabular}[c]{@{}l@{}}
     \specialrule{0em}{0.4pt}{0.4pt}
    \begin{minipage}{\linewidth}
      \includegraphics[width=\linewidth, height=30mm]{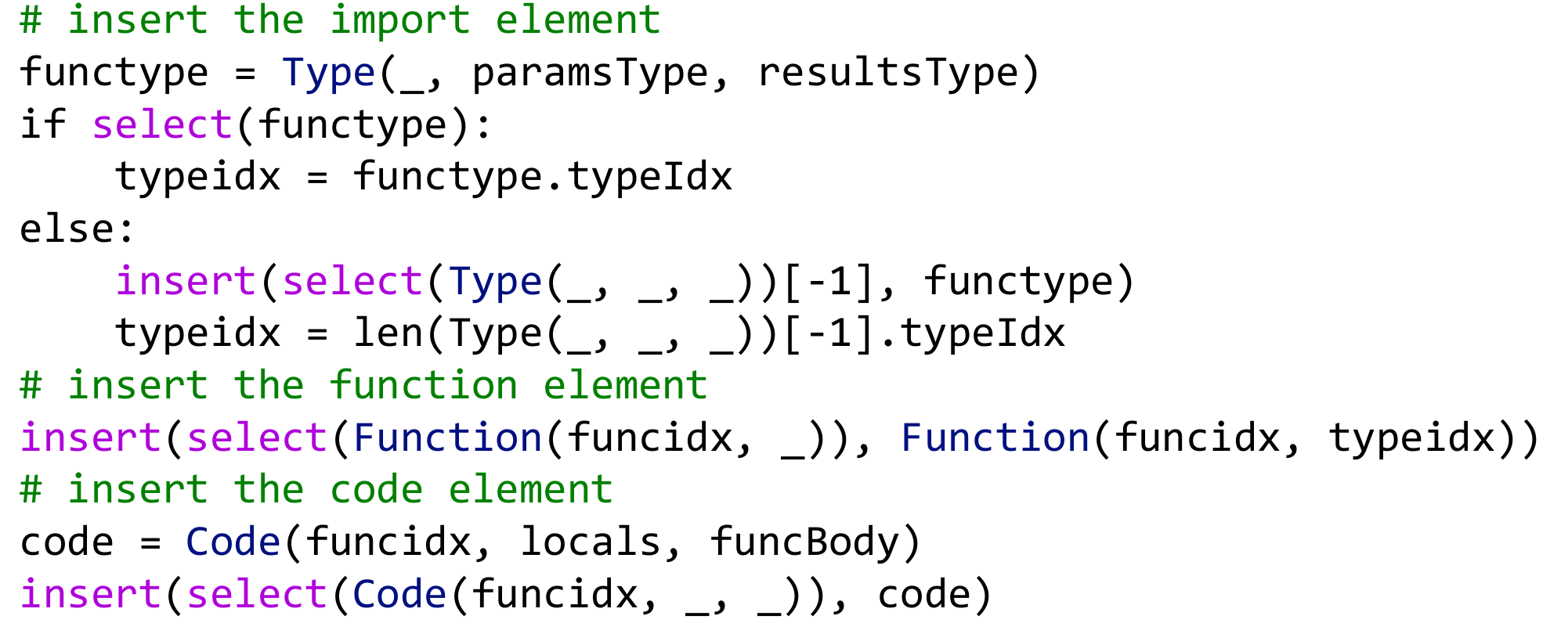}
    \end{minipage}
  \end{tabular} \\ \cline{3-4} 
 &
   &
  \begin{tabular}[c]{@{}l@{}}\code{insertHookFunction}\\ \ \ \ \textit{funcidx}\ :\ \texttt{u32}\\ \ \ \ \textit{hookedFuncIdx}\ :\ \texttt{u32} \\ \ \ \ \textit{funcBody}\ :\ \textit{instruction}*\\ \ \ \ \textit{paramsType}\ :\ (\textbf{\texttt{i32}}$|$\textbf{\texttt{i64}}$|$\textbf{\texttt{f32}}$|$\textbf{\texttt{f64}})*\\ \ \ \ \textit{resultsType}\ :\ (\textbf{\texttt{i32}}$|$\textbf{\texttt{i64}}$|$\textbf{\texttt{f32}}$|$\textbf{\texttt{f64}})*\\ \ \ \ \textit{locals}\ :\ \textit{local}*\end{tabular} &
  \begin{tabular}[c]{@{}l@{}}
     \specialrule{0em}{0.4pt}{0.4pt}
    \begin{minipage}{\linewidth}
      \includegraphics[width=\linewidth, height=37.5mm]{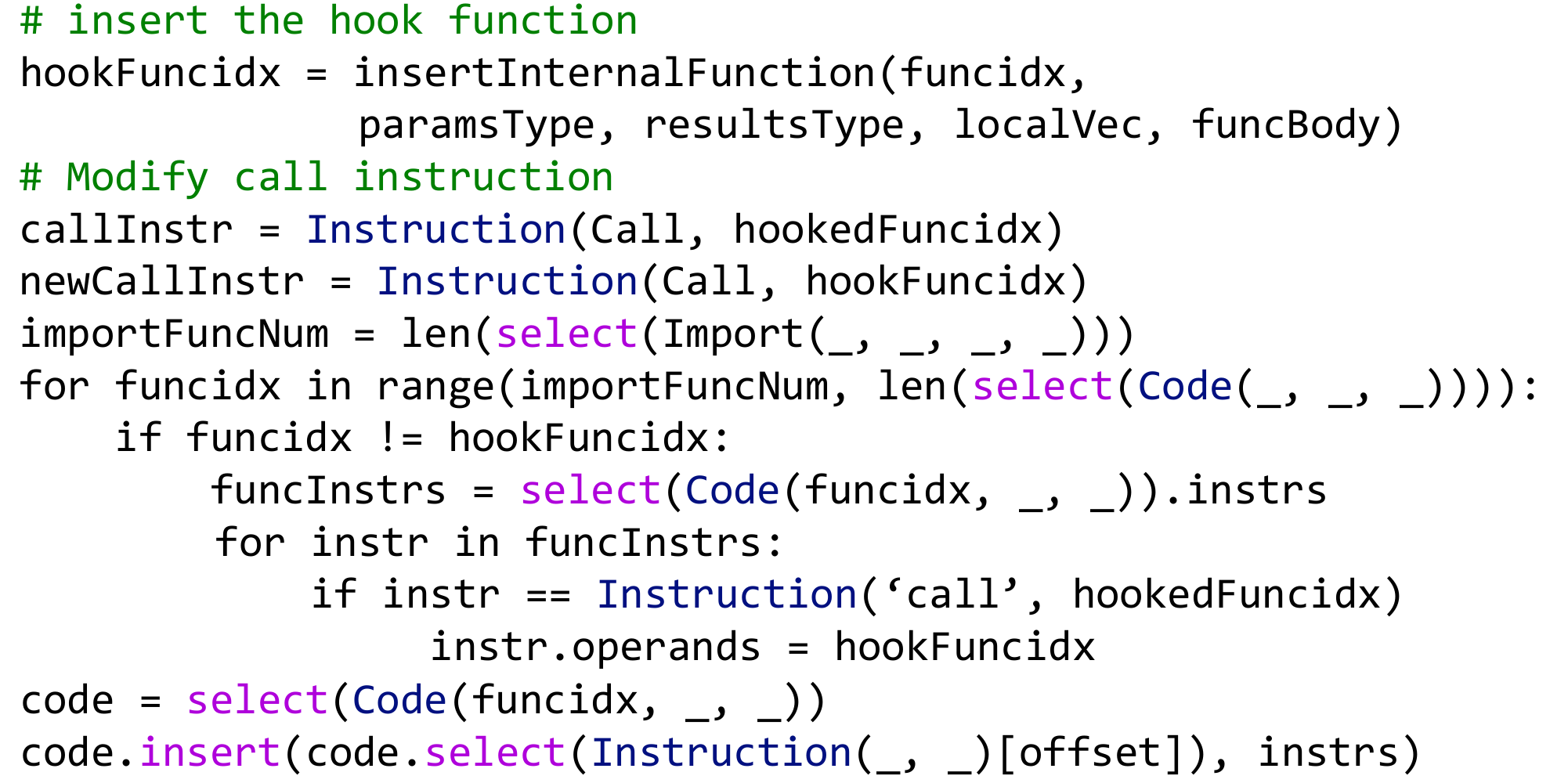}
    \end{minipage}
  \end{tabular} \\ \cline{3-4} 
 &
   &
  \begin{tabular}[c]{@{}l@{}}\code{modifyFunctionInstr}\\ \ \ \ \textit{funcIdx}\ :\ \texttt{u32}\\ \ \ \ \textit{offset}\ :\ \texttt{u32} \\ \ \ \ \textit{instrs}\ :\ \textit{instruction}*\end{tabular} &
  \begin{tabular}[c]{@{}l@{}}
    \specialrule{0em}{0.4pt}{0.4pt}
    \begin{minipage}{\linewidth}
      \includegraphics[width=\linewidth, height=10mm]{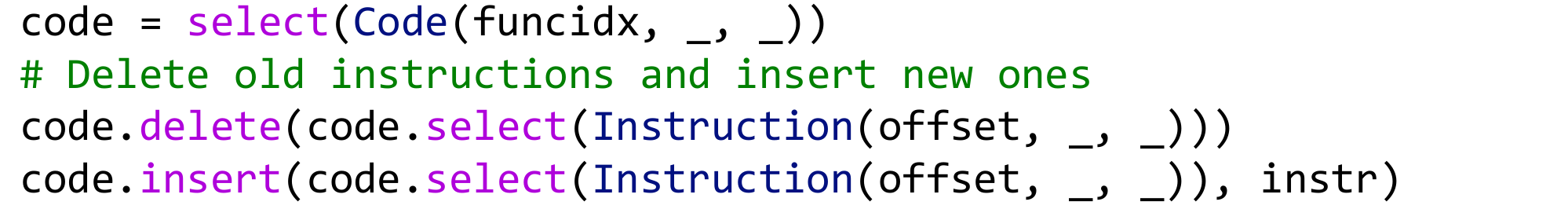}
    \end{minipage}
  \end{tabular} \\ \cline{3-4} 
 &
   &
  \begin{tabular}[c]{@{}l@{}}\code{appendFunctionLocal}\\ \ \ \ \textit{funcidx}\ :\ \texttt{u32} \\ \ \ \ \textit{valType}\ :\ \textbf{\texttt{i32}}$|$\textbf{\texttt{i64}}$|$\textbf{\texttt{f32}}$|$\textbf{\texttt{f64}} \\\end{tabular} &
  \begin{tabular}[c]{@{}l@{}}
    \specialrule{0em}{0.4pt}{0.4pt}
    \begin{minipage}{\linewidth}
      \includegraphics[width=\linewidth, height=5mm]{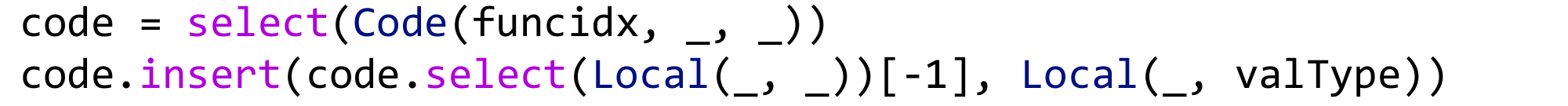}
    \end{minipage}
  \end{tabular} \\ \hline
  
\textbf{\begin{tabular}[c]{@{}l@{}}Custom\\ Content\end{tabular}} &
  Custom &
  \begin{tabular}[c]{@{}l@{}}\code{modifyFunctionName}\\ \ \ \ \textit{funcidx}\ :\ \texttt{u32} \\ \ \ \ \textit{name}\ :\ \texttt{byte}$^*$\end{tabular} &
  \begin{tabular}[c]{@{}l@{}}
    \specialrule{0em}{0.4pt}{0.4pt}
    \begin{minipage}{\linewidth}
      \includegraphics[width=\linewidth, height=5mm]{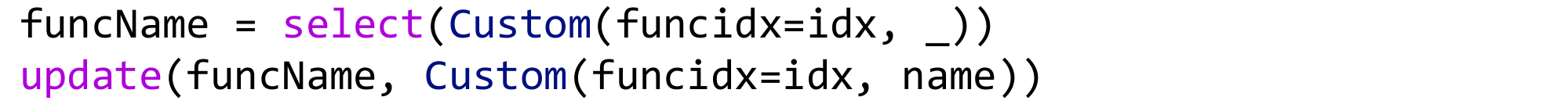}
    \end{minipage}
  \end{tabular}
  \\ \hline
\end{tabular}%
}
\label{table:semantics-rewriter}
\end{table*}

Though we can conclude the flexibility of these four operations through concrete examples, the challenge \textbf{C2} still occurs and has to be addressed.
For example, the $max$ in \textit{memory} elements declares the maximum available space for linear memory. During the rewriting, it is possible to update the $initData$ in a \textit{data} element resulting in exceeding the limitation. 
Another example is that inserting a new or deleting an existing element from any sections may lead to the incontinuity of indices.
Both of these situations invalidate the Wasm binary, which cannot be loaded and executed at all.
To resolve this problem, we implement a fixer that will be automatically invoked after each rewriting requests.
The fixer can determine which section should be fixed, identify the bugs resulting from rewriting requests, and fix them.
Therefore, \textbf{C2} can be addressed after a flexible rewriting process.

\subsection{Semantics Rewriter}
\label{sec:approach:semantics}
Though the section rewriter allows users to rewrite elements and even their fields on a fine-grained level without considering the indices continuity, users still have to make effort to deal with the \textit{section coupling problem} (see \textbf{C3}).
For example, the indices mismatch will occur when indices for type elements are changed but still referred by elements in other sections, like function and code sections.
To this end, users have to manually fix this problem by selecting proper elements in different sections and updating the corresponding fields.

Fixing references on indices between sections, however, is tedious and error-prone.
Moreover, to achieve a little complex functionality, it may be inconvenient for users to solely adopt the fine-grained APIs offered by the section rewriter.
Take appending a function as an example, which is shown in Listing~\ref{lst:insert-function}.

\begin{lstlisting}[caption={Append a function through APIs offered by the section rewriter (rewriting and object initialization APIs are highlighted in magenta and blue respectively).}, label={lst:insert-function}, mathescape=true]
# suppose params, results, locals, and instrs are given by the user
# $\textit{insert}$ the type element
funcType = select(Type(_, params, results))
if funcType:
    typeIdx = funcType[0].typeIdx
else:
    insert(select(Type(_, _, _))[-1], Type(_, params, results))
    typeIdx = select(Type(_, _, _))[-1].typeIdx
# $\textit{insert}$ the function element
insert(select(Function(_, _))[-1], Function(_, typeIdx))
# $\textit{insert}$ the code element
insert(select(Code(_, _, _))[-1], Code(_, local, instrs))
\end{lstlisting}

Suppose the signature of the function ($params$, $results$) as well as its local variables ($local$) and instructions ($instr$) are given by the user.
As we can see, L3 firstly checks if the given signature has been declared. If it is, its index will be kept (L5), or a new index will be calculated by inserting it into the type section (L7 to L8).
Then, the user has to manually link the type index to a function index by inserting a function element (L10).
Finally, the implementation of the function will be appended (L12).
We can see that all newly inserted elements have no concrete $idx$, which is because the fixer we mentioned in \S\ref{sec:approach:section-rewriter} can automatically calculate these indices.

From the instance, we can conclude that rewriting or updating a functionality of a Wasm binary always needs a series of combinations of APIs exposed by the section rewriter.
To ease the burden on users and improve the usability, {\framework} provides another rewriter, named \textit{semantics rewriter}.
We have conducted a comprehensive survey in real-world scenarios on applications (e.g., instrumentation, obfuscation, automatic patch generation, and mutation shown in Fig.~\ref{fig:chaos}) that require binary rewriting as the prerequisite. The survey has covered lots of representative papers~\cite{Lehmann:2021aa,lehmann2019wasabi,SEISMIC,binary-mutate1} and popular repos on GitHub~\cite{wasm-mutate,wasabi}.
Consequently, as shown in Table~\ref{table:semantics-rewriter}, we have abstracted 5 semantics, which cover all 11 sections and offer 31 APIs in total that can be used by these applications.
Specifically, the \textit{global semantics} allows users to arbitrarily update values that can be accessed under the global scope. Through the \textit{import \& export semantics}, users can import or export designated functions. The \textit{memory semantics} can be used to insert another piece of linear memory with a piece of initiated data, while the \textit{function semantics} mainly focuses on updating functions to achieve some goals. Finally, through the \textit{custom semantics}, users can update the debug information to perform obfuscation by changing names of functions. 
For example, the 12-LOC listing~\ref{lst:insert-function} can be abstracted to:
$$\small\texttt{appendInternalFunction(params, results, locals, instrs)}$$

Though it is practical to enhance the usability of implementing some functionalities through calling these 31 APIs, \textbf{C3} has to be handled properly.
For example, inserting new functions may lead to a shifting of indices of existing functions, which may be referred by elements from sections. Such a section coupling problem requires another fixer that investigates and maintains reference relations between elements across sections.
Therefore, after APIs have been invoked, the fixer will be automatically waked to iterate sections and fix reference relations.
Take the \texttt{appendInternalFunction} as an instance, it will examine if any element in \textit{export}, \textit{elem}, and \textit{code} sections has to be fixed.

\section{Implementation \& Evaluation}
\label{sec:imple-eval}
In this section, we evaluate the efficiency, the effectiveness, and the practicability of {\framework} in terms of conducting binary rewriting against representative Wasm binaries.
The research questions and the corresponding experimental results are shown in the following.

\subsection{Implementation}

We have implemented {\framework} with over 4.3K LOC of Python3 code from the scratch.
To avoid reinvent the wheel, some relied modules are based on open-source GitHub projects.
For example, integer literals in Wasm are encoded by LEB128 algorithm~\cite{leb128}. To accelerate the encoding and decoding process, we utilize the highly efficient cyleb128 library~\cite{cyleb128} implemented by Cython.
We have packaged {\framework} into a standard Python library, which can be easily accessed and used by developers.

\subsection{Research Questions \& Experimental Setup}
\label{sec:evaluation:rq}
Our evaluation is driven by the following three research questions:

\begin{itemize}
	\item \textbf{RQ1} Is it efficient to conduct Wasm binaries rewriting through APIs exposed by {\framework}?
	\item \textbf{RQ2} Whether the APIs provided by {\framework} are implemented correctly and effectively?
        \item \textbf{RQ3} Can {\framework} be easily applied to real-world scenarios?
\end{itemize}

To answer these questions, we first selected 10 representative Wasm binaries from WasmBench~\cite{WasmBench}, a well-known micro benchmark that collects tens of thousands of Wasm binaries.
The basic information for them is shown in Table~\ref{table:benchmark}.
We can see that these Wasm binaries are compiled from various programming languages, and cover two typical domains of applying Wasm binaries, i.e., web scripts and standalone applications.
Moreover, they range in size from 3KB to 4MB, which can effectively reflect the ability of {\framework} to handle different Wasm binaries.

\begin{table}[t]
\centering
\caption{Representative Wasm binaries.}
\begin{tabular}{llll}
\toprule
\textbf{Name}      & \textbf{Language} & \textbf{Type} & \textbf{Size (KB)} \\ 
\midrule
zigdom ($B_{1}$)~\cite{wasm-binary1}         & C        & web         & 3      \\
stat ($B_{2}$)~\cite{wasm-binary2}           & C        & standalone  & 45     \\
kindling ($B_{3}$)~\cite{wasm-binary3}       & zig      & web         & 3,373   \\
rustexp ($B_{4}$)~\cite{wasm-binary4}        & Rust     & web         & 935    \\
wasmnes ($B_{5}$)~\cite{wasm-binary5}        & Rust     & web  & 82     \\
base64-cli ($B_{6}$)~\cite{wasm-binary6}     & Rust     & standalone  & 2,415   \\
basic-triangle ($B_{7}$)~\cite{wasm-binary7} & Go       & web         & 1,394   \\
clock ($B_{8}$)~\cite{wasm-binary8}          & Go       & web         & 1,445   \\
go-app ($B_{9}$)~\cite{wasm-binary9}         & Go       & web         & 4,302   \\ 
audio ($B_{10}$)~\cite{wasm-binary10}          & Go       & web  & 8      \\ 
\bottomrule
\end{tabular}
\label{table:benchmark}
\end{table}

All experiments were performed on a server running Ubuntu 22.04 with a 64-core AMD EPYC 7713 CPU and 256GB RAM.

\subsection{RQ1: Efficiency}
\label{sec:rq1}
As we mentioned in \S\ref{sec:approach:semantics}, to make it easier for users to conduct rewriting Wasm binaries, the semantics rewriter exposes a total of 31 APIs, which cover all the legal sections of Wasm.
Therefore, a thorough evaluation of their execution efficiency on Wasm binaries is essential to evaluate {\framework}'s usability.
Based on each Wasm binary in Table~\ref{table:benchmark}, we invoke all 31 APIs with proper arguments and each API is measured by invoking 1000 times.
For example, we call \texttt{insertGlobalVariable} to deliberately insert a global value at the beginning of the global section.

Then, we record how long it will take to finish the designated behavior, including all necessary stages, i.e., parsing, processing, fix-up, and encoding.
The timing results are listed in Table~\ref{table:efficiency}.
It is worth noting that in this experiment each API call requires a parsing process and an encoding process. However, under real scenarios, they are one-shot overhead to complete a given task through calling a series of APIs.
Therefore, we list the parsing and encoding times separately in Table~\ref{table:efficiency} (the second row and the third row, respectively).

\newcommand\multiVI[1]{\multicolumn{10}{c}{#1}}

\begin{table*}[!t]
    \centering
    \caption{Consumed time (\textit{ms}) of invoking APIs provided by the semantics rewriter, as well as the parsing and the encoding time on each Wasm binaries.}
    \resizebox{\textwidth}{!}{%
    \begin{tabular}{r | c c c c c c c c c c}
    \toprule
     & $B_1$ & $B_2$ & $B_3$ & $B_4$ & $B_5$ & $B_6$ & $B_7$ & $B_8$ & $B_9$ & $B_{10}$\\
    \midrule
    \textbf{parsing}        & 6.46  & 55.66  & 173.71  & 1,023.41     & 119.09  & 725.94  & 2,174.82  & 1,794.90    & 6,651.99  & 2.94 \\
    \textbf{encoding}       & 2.58  & 21.28  & 79.29  &  408.89      & 42.32  &  332.95  & 617.66  &  586.59     & 3,294.77  & 2.00 \\
    \midrule
    \textbf{Global Variable} \\
    appendGlobalVariable        & 0.01 & 0.03 & 0.38 & 0.02   & 0.01  & 0.02     & 0.02      & 0.02      & 0.02      & 0.01  \\
    modifyGlobalVariable        & 0.06 & 1.08 & 1.34 & 24.14  & 1.83  & 48.43    & 37.07     & 37.57     & 123.13    & 0.04  \\
    deleteGlobalVariable        & 0.06 & 1.18 & 0.99 & 23.64  & 1.83  & 42.78    & 22.98     & 43.85     & 121.51    & 0.04  \\
    insertGlobalVariable        & 0.39 & 6.84 & 7.72 & 176.34 & 12.26 & 149.25   & 71.87     & 158.33    & 470.11    & 0.20  \\ 
    \midrule        
    \textbf{Import \& Export} \\     
    insertImportFunction        & 0.31 & 4.61 & 4.99 & 76.62  & 9.67  & 101.00   & 48.13     & 102.51    & 300.64    & 0.16  \\
    appendImportFunction        & 0.08 & 1.30 & 1.59 & 29.48  & 2.11  & 38.29    & 15.62     & 38.51     & 118.36    & 0.05  \\
    modifyImportFunction        & 0.07 & 1.35 & 1.10 & 23.30  & 1.79  & 36.41    & 16.58     & 37.37     & 124.53    & 0.05  \\
    deleteImportFunction        & 0.07 & 2.52 & 1.58 & 36.25  & 1.77  & 36.98    & 16.21     & 36.67     & 120.52    & 0.04  \\
    insertExportFunction        & 0.07 & 1.64 & 1.12 & 23.31  & 1.85  & 37.22    & 16.51     & 36.84     & 119.84    & 0.04  \\
    appendExportFunction        & 0.07 & 1.89 & 2.20 & 23.93  & 1.91  & 34.72    & 16.56     & 42.74     & 113.94    & 0.04  \\
    modifyExportFunction        & 0.07 & 1.16 & 0.91 & 22.11  & 1.79  & 34.43    & 15.88     & 36.44     & 132.07    & 0.04  \\
    deleteExportFunction        & 0.07 & 1.29 & 1.85 & 26.81  & 2.13  & 35.22    & 14.97     & 33.75     & 120.78    & 0.04  \\
    \midrule        
    \textbf{Linear Memory} \\     
    appendLinearMemory          & 0.01 & 0.01 & 0.01 & 0.01  & 0.01  & 0.01    & 0.01     & 0.02   & 0.02    & 0.01  \\
    modifyLinearMemory          & 0.01 & 0.02 & 0.01 & 0.03  & 0.02  & 0.02    & 0.03     & 0.02   & 0.02    & 0.01  \\
    \midrule
    \textbf{Function} \\
    insertInternalFunction      & 0.33 & 4.48 & 4.49 & 65.43  & 8.16  & 100.77   & 47.59     & 99.82     & 318.12    & 0.17  \\
    insertIndirectFunction      & 0.38 & 4.64 & 5.87 & 89.97  & 10.11 & 131.80   & 63.39     & 132.27    & 429.44    & 0.21  \\
    insertHookFunction          & 0.74 & 9.88 & 10.87& 176.79 & 20.58 & 260.23   & 119.99    & 261.16    & 833.86    & 0.40  \\
    deleteFuncInstr             & 0.09 & 0.92 & 2.30 & 23.25  & 1.94  & 32.90    & 17.11     & 33.74     & 121.37    & 0.05  \\
    appendFuncInstrs            & 0.07 & 0.86 & 1.32 & 23.55  & 1.81  & 35.42    & 17.90     & 35.13     & 122.17    & 0.04  \\
    insertFuncInstrs            & 0.08 & 0.94 & 1.95 & 24.39  & 1.99  & 33.39    & 16.26     & 33.91     & 126.50    & 0.06  \\
    modifyFuncInstr             & 0.08 & 0.88 & 2.04 & 22.69  & 2.23  & 34.58    & 16.80     & 34.50     & 121.72    & 0.05  \\
    appendFuncLocal             & 0.08 & 0.91 & 1.23 & 23.50  & 2.87  & 33.95    & 23.15     & 33.66     & 118.26    & 0.06  \\  
    \midrule        
    \textbf{Custom Content} \\       
    modifyFuncName              & 0.01 & 0.02 & 0.02 & 0.12   & 0.04  & 0.12     & 0.12      & 0.12      & 0.40      & 0.01  \\
    deleteFuncName              & 0.08 & 0.91 & 1.55 & 23.21  & 1.96  & 34.81    & 15.64     & 33.60     & 121.61    & 0.06  \\
    insertFuncName              & 0.07 & 0.84 & 1.77 & 23.66  & 1.98  & 33.73    & 15.75     & 34.73     & 115.16    & 0.05  \\
    modifyGlobalName            & 0.08 & 0.91 & 1.54 & 22.97  & 2.07  & 34.50    & 15.63     & 34.56     & 120.61    & 0.06  \\
    deleteGlobalName            & 0.07 & 0.84 & 1.14 & 25.27  & 1.84  & 34.71    & 16.14     & 33.63     & 114.15    & 0.05  \\
    insertGlobalName            & 0.08 & 0.94 & 1.05 & 23.53  & 2.81  & 36.37    & 16.06     & 33.16     & 120.71    & 0.06  \\
    insertDataName              & 0.07 & 0.92 & 1.20 & 26.87  & 1.98  & 70.41    & 14.94     & 35.27     & 119.47    & 0.05  \\
    modifyDataName              & 0.08 & 0.89 & 1.39 & 23.40  & 2.45  & 35.04    & 16.10     & 33.45     & 119.95    & 0.05  \\
    deleteDataName              & 0.07 & 0.86 & 3.44 & 23.03  & 2.21  & 35.92    & 15.63     & 34.45     & 126.69    & 0.05  \\
    \bottomrule
\end{tabular}%
}
\label{table:efficiency}
\vspace{-5pt}
\end{table*}

On average, it takes around 1.3s and 0.5s to parse and encode a Wasm binary, respectively.
Although we can see that it took 6.6s to parse $B_9$. However, this is because the Wasm binary consists of more than one million instructions, which will take up more than 100MB of space when converted to WAT format.
Its parsing and encoding times are significantly longer than those of other Wasm binaries. 
Therefore, for most Wasm binaries, we can assume that the parsing and encoding time will not exceed 2 to 3 seconds in total, which is acceptable for users as a one-shot overhead.
Moreover, we can easily observe that each API takes only milliseconds or even less than a millisecond.
Interestingly, among all these APIs, the operations related to \textit{insertion} consume more time than other types of operations.
This is because inserting an entry into a section requires fixing indices of subsequent entries to ensure continuity between indices.
Also, the fixer under the semantics rewriter will have to enumerate all sections to identify if there are mismatched reference relations.
Fortunately, these two fixing processes require no manual intervention.

\begin{tcolorbox}[title= \textbf{RQ-1} Answer, left=2pt, right=2pt, top=2pt, bottom=2pt]
{\framework} has exposed 31 semantics APIs that can efficiently achieve the corresponding goals. Though the parsing and the encoding processes on a Wasm binary take around 1.8 seconds, it is acceptable as a one-shot overhead compared to the negligible time it takes to execute semantics APIs.
\end{tcolorbox}

\subsection{RQ2: Correctness \& Effectiveness}
\label{sec:rq2}
Correctly and effectively achieving the corresponding goals through APIs plays a vital role for {\framework}.
However, it is insufficient to require only the correctness of the implementation of APIs in the section rewriter, which is due to the section coupling problem (see \textbf{C3} in \S\ref{sec:overview:challenges}).
To evaluate the effectiveness of APIs of the semantics rewriter, we pass each Wasm binaries after invoking an API shown in the first column of Table~\ref{table:efficiency} to \textit{wasm-validate}, an official syntax validator.
Then, we manually double checked all 310 (31 APIs * 10 Binaries) cases to make sure the results are inline with the original intents.

According to the results, on the one hand, all rewritten Wasm binaries pass the validation of wasm-validate, indicating valid syntax; on the other hand, all 31 APIs perform correctly in their corresponding functionalities.
Moreover, all 31 APIs resolve the \textbf{C2} and \textbf{C3} correctly.
For example, when the API \texttt{insertImportFunction} is invoked, {\framework} not only rewrites the type section (if necessary) and the import section according to the import function, but also identifies if any instructions are affected, e.g., indices of callee of \texttt{call} instructions and indirect function table in the element section. {\framework} will automatically fix them to keep original semantics intact.
Of course, all these evaluated APIs are passed with valid arguments. If invalid arguments are passed, e.g., inserting a function with a nonexistent index, the underlying \texttt{select} will return an empty list, leading to returning \texttt{false} by the following \texttt{insert} operation (see the formal definitions of \texttt{select} and \texttt{insert} in \S\ref{sec:approach:section-rewriter}).

\begin{tcolorbox}[title= \textbf{RQ-2} Answer, left=2pt, right=2pt, top=2pt, bottom=2pt]
Based on the results of the automated verification tool and manual checks, we can conclude that these semantic APIs perform correctly in both syntax and functionalities.
\end{tcolorbox}
\subsection{RQ3: Practicability}
\label{sec:case-study}
In this section, we will demonstrate the practicability of {\framework} by illustrating some real scenarios that require rewriting Wasm binaries to achieve designated goals.
To ensure the equivalence of reimplementations, we have sent the rewritten Wasm binary to the corresponding tools to examine if identical results can be generated.
The three cases are detailed as follows.

\subsubsection{Case I: Binary Instrumentation}
\label{sec:rq3:case1}
Binary instrumentation is the technique of injecting additional code into the binary file of a program~\cite{binary-instrumentation-concept}. It can be used to collect various runtime information of a program.
D. Lehmann et al.~\cite{lehmann2019wasabi} have implemented a dynamic analysis framework, named Wasabi, whose core is a Wasm binary instrumentation module. It specifies instrumentation rules for each instruction, and the instrumented binary will call pre-defined APIs to obtain runtime information during execution.
For example, to instrument a \texttt{call} instruction, Wasabi inserts two functions (imported through the import section) before and after the \texttt{call}, respectively, to record necessary information.
Through APIs provided by {\framework}, the equivalent functionality can be implemented easily with a few lines of code.

\begin{lstlisting}[caption={Achieve binary instrumentation through APIs provided by {\framework}}, label={lst:binary-instrumentation}, mathescape=true]
instrs = [Instruction("i32.const", 5)]
calleeTypeidx = select(Function(5, _)).typeIdx
calleeType = select(Type(calleeTypeidx, _, _))
# construct call_pre and call_post
callPreFunctype = Type(_, ["i32"] + calleeType.typeArg, calleeType.typeRet)
callPostFunctype = Type(_, calleeType.typeArg, calleeType.typeRet)
# $\textit{insert}$ declaration to import section
appendImportFunction("hooks", "call_pre", callPreFunctype)
callPreFuncidx = select(Import(_, _, _, _))[-1].funcIdx
appendImportFunction("hooks", "call_post", callPostFunctype)
callPostFuncidx = select(Import(_, _, _, _))[-1].funcIdx
# replace the original call instruction
instrs.extend([Instruction("call", callPreFuncidx),
               Instruction("call", 5),
               Instruction("call", callPostFuncidx)])
modifyFuncInstr(Instruction("call", 5), instrs)
\end{lstlisting}

As we can see, listing~\ref{lst:binary-instrumentation} illustrates how to conduct a binary instrumentation around a \texttt{call} instruction, like Wasabi.
Specifically, L1 defines an instruction, \texttt{i32.const 5}, and L2 and L3 retrieve the type of the callee, i.e., the function indexed by \texttt{5}.
Then, L5 and L6 will construct two function signatures according to the callee's type.
L8 invokes an API, named \texttt{appendImportFunction}, to introduce a function, whose name is \texttt{call\_pre} belonging to a module named \texttt{hooks}, into import section.
The same operation is done in L10.
Then, at L13, we will construct a series of instructions, where the original \texttt{call} is wrapped by the newly declared \texttt{call\_pre} and \texttt{call\_post}.
At L16, through another API, the original instruction will be replaced.
Consequently, it is possible for {\framework} to achieve equivalent binary instrumentation like what Wasabi does. 

Due to the fixed pattern of instrumentation used in Wasabi, modifying the underlying code is necessary to achieve other specific binary instrumentation functionalities. In contrast, {\framework} provides a large number of instruction-related general rewriting APIs, making it more flexible and convenient to implement the required instrumentation functionalities. In addition, in~\S\ref{sec:approach:section-rewriter}, our abstraction reduces the complexity of Wasm binaries, lowering the bars and the learning costs of using {\framework}.

\subsubsection{Case II: Software Hardening}
\label{sec:rq3:case2}
\begin{figure}[!t] 
\centering  
\includegraphics[width=0.8\columnwidth]{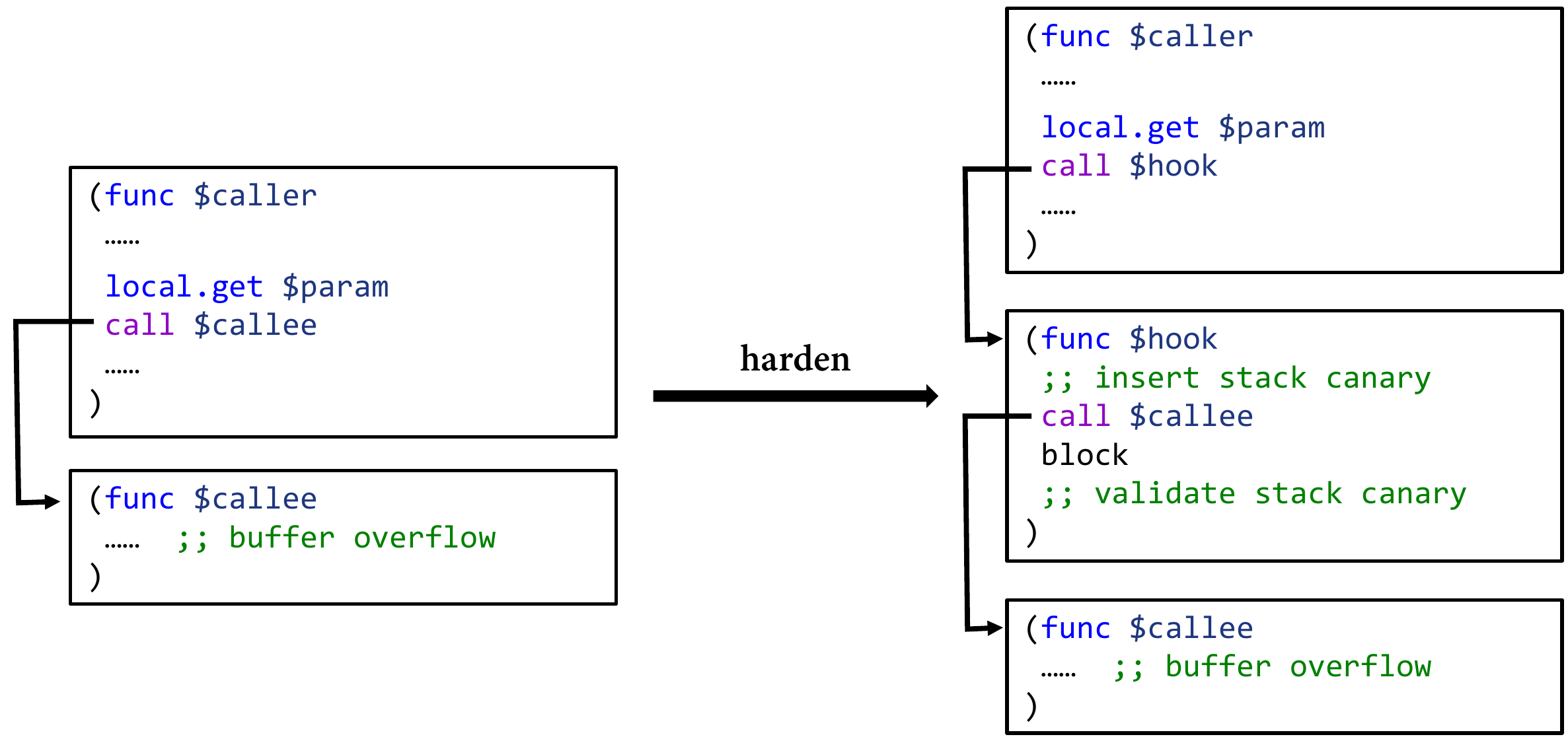} 
\caption{Software hardening through inserting stack canary.} 
\label{fig:stack-canary} 
\end{figure}

Software hardening is to enhance the security and stability of a program by updating it or implementing additional security measures~\cite{software-harden-concept}.
It can also be performed in Wasm binary, which is prone to be affected by kinds of vulnerabilities.
For instance, in Wasm, unmanaged data, like strings, is stored in the linear memory, organized as a stack, and managed by a global variable representing the stack pointer. However, Wasm does not provide stack protection measures, e.g., stack canary~\cite{stack-canary}, to avoid arbitrarily access.
Thus, traditional attacks, like stack overflow, can exploit Wasm binaries leading to out-of-bound read and write~\cite{c-risks-to-wasm}. To mitigate these stack-based attacks, we can conduct Software hardening through {\framework}. Fig.~\ref{fig:stack-canary} shows how to insert a stack canary into a Wasm binary.
As we can see, \texttt{callee} is originally potential for buffer overflow. After the hardening, it will be wrapped by \texttt{hook}, which inserts a stack canary and validates its integrity around the invocation to \texttt{callee}.
Listing~\ref{lst:binary-harden} illustrates how to implement this goal via {\framework}.

\begin{lstlisting}[caption={Achieve Software hardening through APIs provided by {\framework}}, label={lst:binary-harden}, mathescape=true]
# suppose calleeFuncIdx is given by the user
# the first global is a stack pointer
stackPointerIdx = 0
canary = random.randint(1, 10000)
canaryValidateInstrs = [
            Instruction("global.get", stackPointerIdx),
            Instruction("i64.load"),
            Instruction("i64.const", canary),
            Instruction("i64.eq"),
            Instruction("br_if", 0),
            Instruction("unreachable")]
funcbody = [Instruction("global.get", stackPointerIdx),
            Instruction("i32.const", 16),
            Instruction("i32.sub"),
            Instruction("global.set", stackPointerIdx),
            Instruction("global.get", stackPointerIdx),
            Instruction("i64.const", canary),
            Instruction("i64.store", canary)]
calleeTypeidx = select(Function(calleeFuncIdx, _)).typeIdx
calleeType = select(Type(calleeTypeidx, _, _))
for idx, _ in enumerate(calleeType.paramsType):
    funcbody.append(Instruction("local.get", idx))
funcbody.extend([
            Instruction("call", calleeFuncIdx),
            Instruction("block", canaryValidateInstrs),
            Instruction("global.get", stackPointerIdx),
            Instruction("i32.const", 16),
            Instruction("i32.add"),
            Instruction("global.set", stackPointerIdx),
            Instruction("return")])
insertHookFunction(calleeFuncIdx, calleeType.paramsType, calleeType.resultsType, funcbody, $\texttt{locals}$ = [])
\end{lstlisting}

As we can see, at L4, we randomly generate a canary number.
Instructions from L5 to L11, responsible for validating the integrity of canary, are proposed by Fuzzm~\cite{Lehmann:2021aa}.
Similarly, the \texttt{funcbody}, the body of the function \texttt{hook}, defined from L12 to L30 is also proposed by Fuzzm.
To be specific, L12 to L18 deploys the generated canary into the linear memory. From L19 to L22, it dynamically generates the correct number of \texttt{local.get} according to the signature of the callee, which is retrieved by the API \texttt{getFuncFunctype}.
After the instruction \texttt{call \$callee} (L24), we insert the already defined canary validation piece as the operand of a \texttt{block} instruction (L25).
If the value of canary is unchanged, indicating no buffer overflow, the instructions from L26 to L30 will restore the stack and give the control back to \texttt{caller}.

In Fuzzm, stack canary protection is implemented by inserting instructions to write the canary to memory and to verify the canary at the beginning and end of the protected function, respectively. Specifically, since a function may have multiple exit points, fuzzm also modifies the control flow of the function to have only one exit point. Fuzzm's implementation method modifies the instructions and control flow of the protected function, which might disrupt the original semantic of the function. In contrast, {\framework}'s insertHookFunction provides a better solution for this issue, as it minimizes the modifications made to the original function's instructions in the Wasm binary.

\subsubsection{Case III: Fuzzing}
\label{sec:rq3:case3}
Fuzzing is an automated software testing technique that can discover security and stability issues by using random files as input~\cite{fuzzing-concept}.
One of the widely adopted approaches to generate random files is \textit{mutation-based}, i.e., generating new files by mutating existing files.
Wasm-mutate can mutate the given Wasm binary while keeping semantic equivalence.
One of its mutation strategy is \textit{module structure mutation}, e.g., introducing an extra function that has no invoking relationships with existing functions.
This can also be easily done by {\framework}, and a concrete example is shown in Listing~\ref{lst:Wasm-mutate}.

\begin{lstlisting}[caption={Achieve add function mutation through APIs provided by {\framework}}, label={lst:Wasm-mutate}, mathescape=true]
typeSecLen = len(select(Type(_, _, _)))
functype = select(Type(random.randint(0, typeSecLen), _, _))
# construct a function according to a random signature
funcbody = []
for retType in functype.resultsType:
    if retType == "i32":
        funcbody.append(Instruction("i32.const", 0))
    elif retType == "i64":
        funcbody.append(Instruction("i64.const", 0))
    elif retType == "f32":
        funcbody.append(Instruction("f32.const", 0.0))
    elif retType == "f64":
        funcbody.append(Instruction("f64.const", 0.0))
# randomly $\textit{insert}$ the function
importFuncNum = len(select(Import(_, _, _, _)))
internalFuncNum = len(select(Function(_, _)))
funcNum = importFuncNum + internalFuncNum
insertInternalFunction(random.randint(importFuncNum, funcNum), functype.paramType, functype.resultType, funcbody, $\texttt{locals}$ = [])
\end{lstlisting}

As we can see, L1 and L2 randomly select a type serving as a function signature.
According to the designated signature, L4 to L13 construct the body of the function by inserting meaningless instructions to ensure stack balance.
Finally, at L18, the generated function will be randomly inserted into the Wasm binary, and the original semantic keeps intact.

The underlying implementation of wasm-mutate does not include a rewriting component, so all binary rewriting operations involved in the add function mutation strategy require decoding and encoding the relevant sections. In contrast, {\framework} parses the entire wasm binary into an object that we have defined prior before rewriting, which simplifys many extra operations. Additionally, since {\framework} implements the semantics rewriter, inserting a function only requires invoking a single API, without the need to rewrite the relevant sections in turn as in wasm-mutate.

\subsubsection{Other Application Scenarios of {\framework}}
Except for the three concrete applications we mentioned from \S\ref{sec:rq3:case1} to \S\ref{sec:rq3:case3}, {\framework} can also be applied in other scenarios, as summarized in Table~\ref{table:other}.

\begin{table*}[!t]
\centering
\caption{Other application scenarios of {\framework}}
\vspace{-0.1in}
\label{table:other}
\begin{tabular}{lll}
\toprule
\textbf{Scenario}                 & \textbf{Applications}                   & \textbf{Related Semantics APIs}                                                                                    \\ \midrule
Code Obfuscation &
  \begin{tabular}[c]{@{}l@{}}Opaque predicates obfuscation;\\ Memory encryption;\\ Debug info obfuscation\end{tabular} &
  \begin{tabular}[c]{@{}l@{}}\texttt{insertGlobalVariable}\\ \texttt{insertInternalFunction}\\ \texttt{modifyLinearMemory}\\ \texttt{modifyFuncName}\end{tabular} \\ \midrule
Software Testing &
  \begin{tabular}[c]{@{}l@{}}Runtime testing;\\ WASI function testing\end{tabular} &
  \begin{tabular}[c]{@{}l@{}}\texttt{modifyFuncInstr}\\ \texttt{appendImportFunction}\end{tabular} \\ \midrule
Program Repair       &Bug fixing                     & \begin{tabular}[c]{@{}l@{}}\texttt{modifyFuncInstr}\\ \texttt{appendFuncLocal}\end{tabular}                     
\\ \midrule
Software optimization & Instruction optimization   & \begin{tabular}[c]{@{}l@{}}\texttt{modifyFuncInstr}\\ \texttt{appendFuncLocal}\end{tabular}                         \\ \bottomrule
\end{tabular}%
\vspace{-0.1in}
\end{table*}

\noindent \textbf{Code Obfuscation.}
With the help of {\framework}, we can easily implement some traditional code obfuscation methods.
For instance, users can insert global variables or internal functions, as opaque predicates~\cite{opaque}, into a Wasm binary.
Moreover, users can invoke \texttt{modifyLinearMemory} to statically encrypt a piece of linear memory to achieve memory encryption obfuscation. Another function can be inserted and used as a runtime decryptor.
Also, users can change the debug information of the custom section, like obfuscating readable names of functions, to make the Wasm binary unreadable for attackers.

\noindent \textbf{Software Testing.}
The exposed APIs can also be used in software testing.
For example, Y. Zhang et al.~\cite{zhang} have proposed a method to mutate an instruction's operands constantly to examine if the instruction follows the behavior defined in the specification.
Through \texttt{modifyFuncInstr}, users can easily achieve operands rewriting.
Moreover, users can call \texttt{appendExportFunc} to export the result of the instruction to alleviate the workload of comparing the expected result and the actual one.
Consequently, users can apply {\framework} in runtime testing.
The same approach can be applied in testing imported functions as well.

\noindent \textbf{Program Repairing.}
{\framework} can be used to fix bugs in Wasm binaries without source code.
Suppose that an integer overflow occurs in an \texttt{i32.add} instruction. {\framework} can insert a wrapper function around the addition instruction. Within the wrapper, the result of the addition can be verified easily. If an integer overflow happens, users can choose to either correct its result or raise an exception to terminate the execution.

\noindent \textbf{Software Optimization.}
Through APIs of {\framework}, users can conduct instructions optimization.
For example, a piece of Wasm bytecode can be optimized with a higher-level optimization during the compilation, which, however, requires accessing the source code.
To this end, through the APIs we exposed in the section rewriter and the semantics rewriter, users can easily match a piece of code with a pre-defined pattern. Then, through \texttt{modifyFuncInstr} and \texttt{appendFuncLocal}, the code snippet can be updated to an optimized one even without the source code.

\begin{tcolorbox}[title= \textbf{RQ-3} Answer, left=2pt, right=2pt, top=2pt, bottom=1pt]
Our exploration suggests that it is practical to achieve various kinds of complicated Wasm binary rewriting tasks by combing the APIs provided by {\framework}. Comparing with the cumbersome implementation of the specific tasks, the work built on {\framework} is effortless and user-friendly.
\end{tcolorbox}
\section{Related Work}
\noindent \textbf{Wasm Binary Analysis.}
As an emerging stack-based language, WebAssembly can be applied inside or outside the browser~\cite{scnerios}.
Lots of work focused on Wasm binary analysis~\cite{everything,WasmBench,DWARF,minos,slice,eosafe,he2023eunomia}. For example, Lehmann and Pradel~\cite{everything,WasmBench} pay attention to the memory issues in Wasm, e.g., exploitable buffer overflow. They found that some memory issues in the source code will still be exploitable in compiled Wasm binaries.
They also proposed a method that can recover type information by the DWARF debugging information~\cite{DWARF}.
Faraz Naseem et al. proposed MINOS~\cite{minos}, a detector that adopts deep learning to identify unauthorized cryptojacking Wasm binaries.
In addition, Quentin Stiévenart et al.~\cite{slice} overcame the challenges of dependency analysis at the binary level and presented an approach to statically slice Wasm programs to conduct reverse engineering and code comprehension and so on.

\noindent \textbf{Binary Rewriting.}
Binary rewriting can be divided into static and dynamic rewriting, where the former one is focused by this work.
Specifically, static binary rewriting operates on binary files stored in persistent memory~\cite{rewrite-concept}.
Researchers have proposed lots of static rewriting tools against native programs~\cite {Alto,SASI,Diablo,Egalito,McSema}.
Some of them, e.g., Alto~\cite{Alto}, SASI~\cite{SASI} and Diablo~\cite{Diablo}, can only rewrite programs with the assistance of debug information or the ones compiled by a specific compiler.
Recently, E9Patch~\cite{E9Patch} proposes a control-flow-agnostic rewriting method that inserts jumps to trampolines
without the need to move other instructions, which significantly improve the scalability and robustness.
In addition, other tools, e.g., Egalito~\cite{Egalito} and McSema~\cite{McSema}, will lift them into intermediate representations (e.g., LLVM bitcode) and conduct the following rewriting.
The Bytecode Alliance provides two tools, wasm-parser~\cite{wasmparser} and wasm-encoder~\cite{wasm-encoder}, for parsing and encoding Wasm binaries, while using them would face the same challenge of Wasm binary complexity. Thus, we did not rely on them to implement {\framework}. As a comparison, our proposed abstraction model in \S\ref{sec:approach:parser-encoder} could reduce the complexity of Wasm binaries when implementing the section rewriting.
To the best of our knowledge, there is no static or dynamic general binary rewriting framework for Wasm yet.

\section{Limitations \& Discussion}

\noindent
\textbf{Control Flow Rewriting.}
In that work, we mainly focused on the syntactic complexity of Wasm.
However, Wasm also has a special and complicated control flow structure, named \textit{structured control flow}~\cite{control-flow}.
Specifically, the control flow can only go from an internal code block to an external one, and the destination must be either \texttt{block} or \texttt{loop} instruction.
Such an inflexibility makes the control flow rewriting extremely difficult and case-specific.
To the best of our knowledge, there is no work that is able to rewrite the control flow of Wasm in a flexible and general way, which will be one of our future work.

\noindent
\textbf{Scalability Issue of Semantics Rewriter.}
Currently, {\framework} has offered 31 APIs across five semantics that can achieving different functionalities.
Users may concern the scalability, i.e., application scope, of these APIs.
We argue that these APIs are concluded from existing work and several real-world scenarios, which can satisfy most needs.
Moreover, it is practical to implement case-specific semantics APIs through combining four kinds of operations provided by the section rewriter.
The three challenges can be properly handled without users intervention.
On the one hand, as for \textbf{C1}, the Wasm parser can translate the given Wasm binary with complicated syntax into a list of objects, which is intuitive and easy to handle.
On the other hand, the fix against \textbf{C2} and \textbf{C3} will be automatically conducted as we mentioned in \S\ref{sec:approach:section-rewriter} and \S\ref{sec:approach:semantics}, respectively.
Thus, we can conclude the semantics rewriter is scalable and can be widely adopted.

\noindent\textbf{Concerns about Reinventing Wheels.}
Some existing tools have similar design purpose or functionalities with {\framework}, thus there may have concerns about reinventing wheels. We underline that all components in {\framework} are irreplaceable.
Specifically, wasm-parser and wasm-encoder are implemented by the Bytecode Alliance official, however, directly adopting them will overwhelm users.
Though they can parse and encode the given Wasm binary and fully support all sections defined in specification, parsing and encoding require approximately 1,300 and 800 lines of code, respectively. It is tedious and infeasible to compose a script more than 2K LOC to conduct binary rewriting.
In addition, though Wasabi has implemented a wasabi\_wasm module, which is able to conduct static instrumentation, it is insufficient in binary rewriting.
On the one hand, it is designed specifically for binary instrumentation, indicating that it cannot implement removing or updating vectors in sections, which will significantly hinder its ability in terms of binary rewriting.
On the other hand, it does not consider the section coupling challenge as we mentioned in \textbf{C3}. Moreover, some sections are not supported by its module, like the elem section. The lack of ability on rewriting some sections will dramatically impact the flexibility on rewriting.
Consequently, implementing {\framework} instead of combining existing modules or tools does not indicate reinventing wheels.

\section{Conclusion}
In this work, we present {\framework}, a general binary rewriting framework for WebAssembly.
During the rewriting process, {\framework} can properly handle inherent challenges, including the highly complicated binary structure, strict static syntax verification, and coupling among sections.
Based on representative Wasm binaries, varying in source languages, application domains, and even sizes, {\framework} illustrates its efficiency and effectiveness in rewriting process.
Through three cases inspired by existing work and real-world scenarios, {\framework} also proves its practicality and usability.

\bibliography{reference}

\end{document}